\documentclass[a4paper,12pt,oneside,tbtags]{article}
\usepackage{amsmath}
\usepackage{amssymb}
\usepackage{graphics}
\usepackage{epsfig}
\usepackage{color}
\usepackage[nosort]{cite}

\begin{document}

\let\displaymath=\[
\let\enddisplaymath=\]
\def\[{\begin{equation}}
\def\]{\end{equation}}
\def\<{\begin{myeqnarray}}
\def\>{\end{myeqnarray}}


\def\da{\dot{a}}
\def\db{\dot{b}}
\def\G{\Gamma}
\def\D{\Delta}
\def\L{\Lambda}
\def\S{\Sigma}
\def\a{\alpha}
\def\b{\beta}
\def\g{\gamma}
\def\d{\delta}
\def\e{\varepsilon}
\def\m{\mu}
\def\tr{\mbox{tr}}
\def\n{\nu}
\def\s{\sigma}
\def\r{\rho}
\def\l{\lambda}
\def\t{\tau}
\def\o{\omega}
\def\O{\Omega}
\def\v{\varrho}
\def\vt{\vartheta}
\def\mc{\mathcal}
\def\N{\nabla}
\def\p{\partial}
\def\la{\langle}
\def\ra{\rangle}
\def\dg{\dagger}
\def\wt{\widetilde}
\def\Cop{\bbbc}
\def\Zop{\bbbz}
\def\Rop{\bbbr}
\def\Nop{\bbbn}
\def\bbbz {{\sf Z\!\!Z}}
\def\bbbp {{\sf P\!\!|}}
\def\bbbr {{\rm I\!R}}
\def\bbbn {{\rm I\!N}}
\def\RR{R-R }

\makeatletter
\let\old@makecaption=\@makecaption
\def\@makecaption{\small\old@makecaption}
\makeatother

\newcommand{\hypref}[2]{\ifx\href\asklfhas #2\else\href{#1}{#2}\fi}
\newcommand{\remark}[1]{\textbf{#1}}
\newcommand{\figref}[1]{fig. \ref{#1}}
\newcommand{\Figref}[1]{Fig. \ref{#1}}
\newcommand{\tabref}[1]{tab. \ref{#1}}
\newcommand{\indup}[1]{_{\mathrm{#1}}}
\newcommand{\tsum}{{\textstyle\sum}}
\newcommand{\tprod}{{\textstyle\prod}}
\newcommand{\bu}{$\bullet$ \,}

\newcommand{\sfrac}[2]{{\textstyle\frac{#1}{#2}}}
\newcommand{\half}{\sfrac{1}{2}}
\newcommand{\quarter}{\sfrac{1}{4}}

\newcommand{\Op}{\mathcal{O}}
\newcommand{\order}[1]{\mathcal{O}(#1)}
\newcommand{\eps}{\varepsilon}
\newcommand{\Lagr}{\mathcal{L}}
\newcommand{\superN}{\mathcal{N}}
\newcommand{\gym}{g_{\scriptscriptstyle\mathrm{YM}}}
\newcommand{\gtwo}{g_2}
\newcommand{\Tr}{\mathop{\mathrm{Tr}}}
\renewcommand{\Re}{\mathop{\mathrm{Re}}}
\renewcommand{\Im}{\mathop{\mathrm{Im}}}
\newcommand{\Li}{\mathop{\mathrm{Li}}\nolimits}
\newcommand{\cdott}{\mathord{\cdot}}
\newcommand{\singlet}{{\mathbf{1}}}

\newcommand{\lrbrk}[1]{\left(#1\right)}
\newcommand{\bigbrk}[1]{\bigl(#1\bigr)}
\newcommand{\vev}[1]{\langle#1\rangle}
\newcommand{\normord}[1]{\mathopen{:}#1\mathclose{:}}
\newcommand{\lrvev}[1]{\left\langle#1\right\rangle}
\newcommand{\bigvev}[1]{\bigl\langle#1\bigr\rangle}
\newcommand{\bigcomm}[2]{\big[#1,#2\big]}
\newcommand{\lrabs}[1]{\left|#1\right|}
\newcommand{\abs}[1]{|#1|}

\newcommand{\nn}{\nonumber}
\newcommand{\nln}{\nonumber\\}
\newcommand{\nl}{\nonumber\\&&\mathord{}}
\newcommand{\nle}{\nonumber\\&=&\mathrel{}}
\newcommand{\eq}{\mathrel{}&=&\mathrel{}}
\newenvironment{myeqnarray}{\arraycolsep0pt\begin{eqnarray}}{\end{eqnarray}\ignorespacesafterend}
\newenvironment{myeqnarray*}{\arraycolsep0pt\begin{eqnarray*}}{\end{eqnarray*}\ignorespacesafterend}

\newcommand{\NPB}[3]{{\it Nucl.\ Phys.} {\bf B#1} (#2) #3}
\newcommand{\CMP}[3]{{\it Commun.\ Math.\ Phys.} {\bf #1} (#2) #3}
\newcommand{\PRD}[3]{{\it Phys.\ Rev.} {\bf D#1} (#2) #3}
\newcommand{\PLB}[3]{{\it Phys.\ Lett.} {\bf B#1} (#2) #3}
\newcommand{\JHEP}[3]{{JHEP} {\bf #1} (#2) #3}
\newcommand{\hepth}[1]{{\tt hep-th/#1}}
\newcommand{\ft}[2]{{\textstyle\frac{#1}{#2}}}
\newcommand{\cO}{{\cal O}}
\newcommand{\cT}{{\cal T}}
\def\ss{\scriptstyle}
\def\st{\scriptstyle}
\def\sst{\scriptscriptstyle}
\def\ra{\rightarrow}
\def\lra{\longrightarrow}
\newcommand\Zb{\bar Z}
\newcommand\bZ{\bar Z}
\newcommand\bF{\bar \Phi}
\newcommand\bP{\bar \Psi}
\newcommand\hint{h_{int}}
\newcommand\hep[1]{[#1]}

\renewcommand{\theequation}{\thesection.\arabic{equation}}

\thispagestyle{empty}
\begin{flushright}
{\sc\footnotesize hep-th/0402211}\\
{\sc AEI 2004-019}
\end{flushright}
\vspace{1cm}
\setcounter{footnote}{0}
\begin{center}
{\Large{\bf Decay widths of three-impurity states in the BMN correspondence\par}
    }\vspace{10mm}
{\sc Petra Gutjahr and  
Jan Plefka\\[7mm]
Max-Planck-Institut f\"ur Gravitationsphysik\\
Albert-Einstein-Institut\\
Am M\"uhlenberg 1, D-14476 Potsdam, Germany}\\ [2mm]
{\tt petra.gutjahr,jan.plefka@aei.mpg.de}\\[20mm]

{\sc Abstract}\\[2mm]
\end{center}
We extend the study of the quantum mechanics of BMN gauge theory to
the sector of three scalar impurities at one loop and all genus. The
relevant matrix elements of the non-planar one loop dilatation
operator are computed in the gauge theory basis. After a similarity
transform the BMN gauge theory prediction for the corresponding piece
of the plane wave string Hamiltonian is derived and shown to agree
with light-cone string field theory. In the three-impurity sector
single string states are unstable for the decay into two-string states
at leading order in $g_2$. The corresponding decay widths are
computed.

\newpage
\setcounter{page}{1}

\section{Introduction}

The duality of superstrings in a maximally supersymmetric plane-wave
background and $U(N)$ ${\cal N}=4$ Super Yang-Mills 
theory in a particular scaling limit
proposed by Berenstein, Maldacena and Nastase \cite{BMN} has been
the subject of intense study over the past two years\footnote{For reviews 
of this subject see \cite{reviews}.}. This duality may be viewed as a ``corollary'' to
the AdS/CFT correspondence \cite{malda}, as it arises through
a Penrose limit 
on the supergravity background \cite{Blau:2001ne}  
and, in parallel, through a double scaling large $N$,
large $U(1)_R$ charge limit on the gauge theory side \cite{us1,Boston7}. 
The remarkable feature of this novel correspondence is its seemingly perturbative 
structure, allowing for dynamical, quantitative tests 
extending into the true stringy domain of higher-level excitations 
\cite{Metsaev} as well as string interactions \cite{SFT}.

At the ``heart'' of the BMN correspondence lies the identification
$E_{\rm l.c.}/ \mu =\Delta - J$, relating the light-cone energy of string excitations to the
scaling dimensions $\Delta$ and the $U(1)_R$ charge $J$ 
of the suitably chosen dual gauge theory 
operators, the so called ``BMN operators''.
Lifted to the level of interacting strings and non-planar 
Yang-Mills theory the current understanding of the duality
states that \cite{oprel}
\[
\frac{{\cal H}_{\rm l.c.}}{\mu} \mathrel{\hat =} {\cal D}- J\cdot \mathbf{1}
\label{HREL}
\]
relating the interacting string field theory Hamiltonian ${\cal H}_{\rm l.c.}$
to the dilatation operator ${\cal D}$ of ${\cal N}=4$ Super Yang-Mills. Here
$\mu$ is the mass parameter of the plane wave background space-time. 
The perturbative expansion of ${\cal D}$ is controlled 
by the effective coupling constant $\lambda'=\frac{g_{\rm YM}^2\, N}{J^2}$
and the genus counting parameter $g_2=\frac{J^2}{N}$ and one works
in the limit $N,J\to\infty$ and $J^2/N$ finite \cite{us1,Boston7,us2,QM}.
A large number of tests of the operator identification \eqref{HREL}
has been reported (see \cite{reviews} and references therein). However, most of these tests have been restricted
to the sector of two-impurity BMN operators, or equivalently to level two excitations
of the plane wave superstring. In this paper we will push the
analysis to the level of three-impurities, corresponding to
string excitations of level three
\begin{align}\label{stringsqr}
 &\a_{n_1}^{I\, \dagger} \a_{n_2}^{J\, \dagger} \tilde{\a}_{n_3}^{K\,\dagger}
|0,p^+\rangle, & & \frac{E_{lc}}{\m}=\sum_{i=1}^{3}\sqrt{1+n_i^2 \l'} \qquad\text{with} 
\quad n_1+n_2+n_3=0 \, .
\end{align}
Studies of higher-impurity interactions have been undertaken in 
\cite{planar3imp} and \cite{gomis}. In \cite{planar3imp} an alternative
proposition for the
duality relation of \eqref{HREL} based on gauge-theory three-point functions 
was tested, involving three-scalar impurities. In \cite{gomis} a 
direct correspondence between Feynman diagrammatic calculations
in gauge-theory two-point functions and string field theory calculations
for any number of impurities was observed.

In our work, we shall study the decay of three-impurity states into the continuum of 
degenerate two-string states using
the efficient reformulation of BMN gauge theory in terms of a
quantum mechanical system \cite{QM}.
As observed in \cite{GF} such an instability for decay also exists
in the two-impurity sector, there it is, however, suppressed at leading
order in $g_2$, i.e. the leading order decay is
of single string states into degenerate triple-string states. In the three-impurity sector
(and for more impurities \cite{Bonderson}) a non-vanishing decay rate occurs 
already at leading order in $g_2$. Similar decay rates were computed in the
plane-wave limit of ``little''-string theory in \cite{elias}.

The standard way
to find the scaling dimensions $\Delta_\alpha$ of a set of
conformal fields ${\cal O}_\alpha$ in a conformal field theory
is to compute the two
point functions, whose form is determined by conformal symmetry to be
\begin{equation} \label{2-point-fct}     
\langle {\mathcal O}_{\a}(x) \bar{{\mathcal O}}_{\b}(y) \rangle = 
\frac{\d_{\a \b}}{(x-y)^{2 \D_{{\mathcal O}_{\a}}}} \, ,
\end{equation}
and to deduce the associated scaling dimensions
$\Delta_\alpha$. In practice, however, 
it is extremely laborious to diagonalize the two-point functions
starting from a (natural) basis of gauge invariant operators
due to the effect of operator mixing.
A very efficient method to compute the scaling dimensions
in perturbative gauge theory was introduced in \cite{QM} (see also \cite{Janik}) and further 
refined in \cite{DilOp}. The idea is to shift attention away from the
explicit two-point function toward the dilatation operator ${\cal D}$ acting
on fields at the origin, whose eigenvalues $\Delta_\alpha$
are the scaling dimensions
\[
{\cal D}\circ {\cal O}_\alpha = \Delta_\alpha\, {\cal O}_\alpha\, .
\]
The dilatation operator can be constructed in perturbation theory. 
Up to one quantum loop order and in the sector of pure scalar ${\cal N}=4$
Super Yang-Mills  operators
($\phi_I$, $i=1,\ldots,6$)
the dilatation operator takes the simple form 
\cite{us1,us2}\footnote{As a matter of fact ${\cal D}$ is known for all excitations of ${\cal N}=4$
Super Yang-Mills at one loop order \cite{niklas1} and in the subsector 
of two complex scalars at two- and three-loop order \cite{DilOp,nkp}.}
\<
\lefteqn{{\cal D}= \Tr(\phi_I\,\check\phi_I) }\\
&&\qquad- \frac{g_{\rm YM}^2}{16\pi^2}\, \left (
:\Tr[\phi_I,\phi_J]\, [\check\phi_I,\check\phi_J]: +\ft 1 2
:\Tr[\phi_I,\check\phi_J]\, [\phi_I,\check\phi_J]: 
\right ) +{\cal O}(g_{\rm YM}^4) \nn
\>
where $:\,\,:$ denotes normal ordering and $\check\phi_I$ is the matrix derivative
\[
(\check\phi_I)_{ab}=\frac{\delta}{\delta{(\phi_I)}_{ba}}\qquad a,b=1,\ldots,N\, .
\]
In the BMN limit one considers the complexified scalar field $Z=\ft{1}{\sqrt 2}(\phi_5+i\phi_6)$
carrying unit $U(1)_R$ charge and restricts to the subsector of operators
of total charge $J$. The action of  $({\cal D} - J\cdot \mathbf{1})$
on BMN operators is thus given by the number of impurity
insertions ($\phi_1,\ldots,\phi_4$) into the string of $Z$'s
 plus the one (and higher) loop pieces of ${\cal D}$. 
The large $J$ limit then serves as a continuum limit of the (discrete)
action of $({\cal D} - J\cdot \mathbf{1})$ on BMN operators, yielding an effective
quantum mechanical system describing BMN gauge theory \cite{QM}. The resulting
Hamiltonian $H$ consists of a free piece and an interacting part of order
$g_2$ responsible for string splitting and joining processes
\[
H=H_0+g_2 (H_++H_-):=\lim_{\stackrel{N,J\to \infty,}{{}_{N/J^2\, \text{fixed}}}} 
\left ( {\cal D}- J\cdot \mathbf{1}
\right ) \,.
\label{Hamiltonian}
\]
Hence, a string field theory Hamiltonian emerges from the gauge theory and 
terminates (at ${\cal O}(\lambda')$) with order $g_2$ terms. 

However, the Hamiltonian emerging from the gauge theory cannot be immediately
compared to the Hamiltonian arising in a light-cone string field theory
treatment of the problem \cite{SFT,HSSV}. This is due to the fact that $H$ is not Hermitian,
whereas the string field theory Hamiltonian is.

The problem can be understood as follows:
In the quantum mechanical system the
inner product of states is identified with the {\em planar} part of the
{\em free} theory two point functions
\[
\langle a|b\rangle := \langle {\cal O}_a(x) \, \bar{\cal O}_b(y)\rangle_{0,
{\rm planar}}\cdot
|x-y|^{\Delta_a^{(0)}}\, ,
\]
where the last factor on the right-hand side strips off all
space dependencies and the index ``$0$'' denotes correlation functions in the
free theory. As a matter of fact $H$ is {\em not}
Hermitian with respect to $\langle\, |\, \rangle$, but with respect to $\langle\,
|\, \rangle_{g_2}$, which is the inner product induced by the {\em full non
planar, free} two-point function. One then defines a Hermitian (w.r.t. $\langle\, |\,\rangle$)
operator $S$
by
\begin{equation} \label{rel_inner_product}
\langle a |b \rangle_{g_2}=\langle a |S|b \rangle := 
\langle {\cal O}_a(x) \, \bar{\cal O}_b(y)\rangle_{0, {\rm full}}\cdot
|x-y|^{\Delta_a^{(0)}}\, ,     
\end{equation}
and therefore has
\begin{equation}
S \, H = H^{\dg} \, S\,.       
\end{equation}
Hence the Hamiltonian $H$ is only quasi Hermitian. However, there is a
natural basis $|\tilde a\rangle$, related to $|a\rangle$ through 
the non-unitary transformation
\begin{equation} \label{stringbasis}
|\tilde{a}\rangle = S^{-1/2} |a\rangle \;,
\end{equation}
which diagonalizes $\langle \, | \, \rangle_{g_2}$.
One then defines a new Hermitian operator $\tilde H$  through
\begin{equation}\label{h_tilde}
\langle \tilde{a} | H | \tilde {b}\rangle_{g_2} = \langle a 
|S^{-1/2} H S^{-1/2}|b \rangle_{g_2} = \langle a |S^{1/2} H 
S^{-1/2}|b \rangle = \langle a |\tilde{H}|b \rangle\; ,
\end{equation}
known as the string Hamiltonian.  If we identify a BMN operator with
$k$ traces with the state $|k\rangle$, $|\tilde{k}\rangle$ corresponds
to a string state with $k$ strings and hence one usually refers to the
basis
\eqref{stringbasis} as the string basis.
The matrix elements of $\tilde H$ should
match those obtained in light-cone string field theory -- up to a 
possible unitary transformation. We would like to stress that, also from
a purely gauge theoretic perspective, the Hermitization of $H$ through
conjugation with $S^{-1/2}$ is a very natural construction.

In the first part of this paper we will calculate
the matrix elements of $\tilde{H}$ up to one-loop and first order in
$g_2$ in the three-impurity sector. 
They are given by (see \eqref {h_tilde})
\begin{equation}
\langle a |\tilde{H}|b\rangle = 
\langle a|\big(1+\tfrac{1}{2}g_2 \S \big)\, H \, \big(1-\tfrac{1}{2}g_2 \S 
\big)| b\rangle
+ {\mathcal O}(g_2^2) \;,  
\end{equation}
where $\S$ denotes the $g_2$ contribution to $S$. Thus we need to
evaluate the
matrix elements for $H$ (section 2.1) and $\S$ (section 2.2). 
In section 2.3 the results for the string Hamiltonian
$\tilde{H}$ are presented and shown to agree with string field theory
in section 3. In section 4 we finally evaluate the decay widths for the
transition of a single-string state into the continuum of degenerate
double-string states.
     

\section{The gauge theory computation}
\setcounter{equation}{0}

\subsection{Matrix elements of the effective 1-loop vertex operator}

We shall be  interested in three-impurity BMN operators of total $U(1)_R$ 
charge $J$ and  three (different)
scalar impurity insertions $\phi_i$ with $i=1,2,3,4$. There are three distinct
ways in which these impurities can be distributed over separate traces. 
If all impurities fall in a single trace one has
\[ \label{o(3)}
{\mathcal O}^{\text{\tiny 123}}_{p_1,p_2,p_3}{\mathcal O}_{J_1}
\ldots{\mathcal O}_{J_l} = 
{\rm Tr} \big[ \phi_1 \, Z^{p_1} \, \phi_2 \, Z^{p_2} \, \phi_3 \, 
Z^{p_3}\big]{\rm Tr} \left[ Z^{J_1} \right]\ldots {\rm Tr} \left[ Z^{J_l} \right]
\]
with $p_1 + p_2 + p_3 = J_0$ and $\sum_{k=0}^{l}J_k=J$ and
$Z=\ft {1}{\sqrt 2} (\phi_5+i\phi_6)$.
When acting with the Hamiltonian \eqref{Hamiltonian} on this operator
one obtains, in a schematic notation,
\begin{align} \label{h_o(3)}
&H_0 \circ {\mathcal O}_{3; l} \propto {\mathcal O}_{3; l}\notag\\
&H_- \circ {\mathcal O}_{3; l} \propto {\mathcal O}_{3; l-1}\notag\\
&H_+ \circ {\mathcal O}_{3; l} \propto {\mathcal O}_{3; l+1} + {\mathcal O}_{2,1; l}\;.
\end{align}
where ${\cal O}_{a_1,\ldots,a_j;l}$ denotes an operator with $a_i$ impurities
in the $i$th trace and $l$ gives the number traces without impurities.
$H_0$ conserves the total number of traces, while in $H_-$ ($H_+$) one trace 
is removed (added). The total number of $Z$'s never changes.
Contributions from $H_+$, which redistribute the impurities among several 
traces, did not occur for the case of two scalar impurities \cite{QM}. 
Therefore we need to extend our basis of three impurity operators with 
\begin{align}
\label{o(2,1)} 
{\mathcal O}^{\text{\tiny ab}}_{p_a,p_b}{\mathcal O}^{\text{\tiny c}}_{p_c}
{\mathcal O}_{J_1}\ldots{\mathcal O}_{J_l} &
= {\rm Tr} \left[ \phi_a \, Z^{p_a} \, \phi_b \, Z^{p_b} \right]{\rm Tr} 
\left[ \phi_c \, Z^{p_c} \right]\left[ Z^{J_1} \right]\ldots {\rm Tr} 
\left[ Z^{J_l} \right] \nonumber \\
&\mbox{where} \quad (a,b)(c)\in\{(1,2)(3);(2,3)(1);(1,3)(2)\}
\end{align}
and
\[
{\mathcal O}^{\text{\tiny 1}}_{p_1}{\mathcal O}^{\text{\tiny 2}}_{p_2}
{\mathcal O}^{\text{\tiny 3}}_{p_3}{\mathcal O}_{J_1}\ldots{\mathcal O}_{J_l} 
= {\rm Tr} \left[ \phi_1 \, Z^{p_1} \right]{\rm Tr} 
\left[ \phi_2 \, Z^{p_2} \right]{\rm Tr} \left[ \phi_3 \, Z^{p_3} \right]
{\rm Tr} \left[ Z^{J_1} \right]\ldots {\rm Tr} \left[ Z^{J_l} \right] \, .
\]
The action of $H$ on these two classes of operators reads
\begin{align} \label{h_o(2,1)}
H_0 \circ {\mathcal O}_{2,1; l} &\propto {\mathcal O}_{2,1; l}			      
& H_0 \circ {\mathcal O}_{1,1,1; l} &= 0 \notag\\
H_- \circ {\mathcal O}_{2,1; l} &\propto {\mathcal O}_{2,1; l-1}+ {\mathcal O}_{3; l}\qquad\quad\text{and} \quad
& H_- \circ {\mathcal O}_{1,1,1; l} &\propto  {\mathcal O}_{2,1; l} \notag\\
H_+ \circ {\mathcal O}_{2,1; l} &\propto {\mathcal O}_{2,1; l+1}		      
& H_+ \circ {\mathcal O}_{1,1,1; l} &= 0\;. 
\end{align}
The explicit expressions of \eqref{h_o(3)} and \eqref{h_o(2,1)} can be found 
in Appendix \ref{appH_0}. 

In
the BMN limit ($N,J\to \infty$ with $J^2/N$ fixed) one introduces
continuum variables. The operators above are then replaced by
continuum states
\begin{align}
{\mathcal O}^{\text{\tiny 123}}_{p_1,p_2,p_3}{\mathcal O}_{J_1}\ldots{\mathcal O}_{J_l} 
&\rightarrow \;\frac{\sqrt{N^{J+3}}}{J} \;|x_1,x_2,x_3 
\rangle^{\text{\tiny 123}}|r_1\rangle \ldots |r_l\rangle\\
{\mathcal O}^{\text{\tiny ab}}_{p_a,p_b}{\mathcal O}^{\text{\tiny c}}_{p_c}
{\mathcal O}_{J_1}\ldots{\mathcal O}_{J_l} 
&\rightarrow \;\frac{\sqrt{N^{J+3}}}{J} \;|x_a,x_b\rangle^{\text{\tiny ab}}
|x_c\rangle^{\text{\tiny c}}|r_1\rangle \ldots |r_l\rangle\\
{\mathcal O}^{\text{\tiny 1}}_{p_1}{\mathcal O}^{\text{\tiny 2}}_{p_2}
{\mathcal O}^{\text{\tiny 3}}_{p_3}{\mathcal O}_{J_1}\ldots{\mathcal O}_{J_l} 
&\rightarrow \;\frac{\sqrt{N^{J+3}}}{J} \;|x_1\rangle^{\text{\tiny 1}}
|x_2\rangle^{\text{\tiny 2}}|x_3\rangle^{\text{\tiny 3}}|r_1\rangle \ldots |r_l\rangle
\end{align}
with $x_j := \frac{p_j}{J}$; $r_k:= \frac{J_k}{J}; \, k= 0, \ldots l$;
$x_j,r_k \in \left[0, 1 \right]; \,x_1 + x_2 + x_3 = r_0$ and
$\sum_{k=0}^{l}r_k=1$. 

In order to understand the prefactors, let us have a look
at the planar tree-level two-point function of two operators of the
type ${\mathcal O}_{3;l}$ (the result is the same for ${\mathcal
O}_{2,1;l}$ and ${\mathcal O}_{1,1,1;l}$)
\begin{align}
\lefteqn{\langle {\mathcal O}^{\text{\tiny 123}}_{p_1,p_2,p_3}
{\mathcal O}_{J_1}\ldots{\mathcal O}_{J_l} 
\bar{{\mathcal O}}^{\text{\tiny 123}}_{q_1,q_2,q_3}
\bar{{\mathcal O}}_{K_1}\ldots\bar{{\mathcal O}}_{K_{l'}}\rangle_{0,{\rm planar}}=}\nn\\
&\qquad\qquad N^{J+3} \delta_{p_1,q_1}\delta_{p_2,q_2}\delta_{p_3,q_3}\delta_{l,l'}
\sum_{\pi \in S_l} \prod_{k=1}^l \delta_{J_k,K_{\pi(k)}} J_k\;,
\end{align}
which will be identified with the inner product of the corresponding
states.  The factor of $N^{J+3}$ is absorbed by the definition of the
states. As to the powers of $J$ note that
 we are going to deal only with operators having the
same number of $Z$'s. Hence we can express $p_3$ by $p_3 = J_0 - p_1 - p_2$
and analogously $q_3$ by $q_3 = K_0 - q_1 - q_2$ with $J =
\sum_{k=0}^{l} J_k = \sum_{k=0}^{l'}K_k$ and thus $\delta_{p_3,q_3}$
becomes redundant. The remaining $p$'s have to obey the
condition $0 \le p_1+p_2 \le J_0$ affecting the limits of summation
(or integration for the $x$'s) later on. On the other hand, we would
like to compare our results to those of string field theory, where the
inner product of two states is given by $\langle i|j \rangle \propto
\delta(p_i^+-p_j^+)$ with $p^+$ being the light-cone momentum of a
state. Therefore we need a factor of $J^{m+2}$ to convert each
Kronecker-delta into a $\delta$-function. One finds that replacing
$J_k$ by $r_k \;J$ gives $J^l$ and thus only an additional factor of
$J^2$ is required. The inner products of the continuum states are then
given by
\begin{align}\label{ips}
\langle r_1| \ldots \langle r_l| \:^{\text{\tiny 123}}
\langle x_1,x_2,x_3|x_1',x_2',x_3'\rangle^{\text{\tiny 123}}\:|r_1'
\rangle \ldots |r_{l'}' \rangle& = \notag\\
\langle r_1| \ldots \langle r_l|\: ^{\text{\tiny c}}\langle x_c|\: ^{\text{\tiny ab}}
\langle x_a,x_b|x_a',x_b'\rangle^{\text{\tiny ab}}\:|x_c'\rangle^{\text{\tiny c}}\:
|r_1'\rangle \ldots |r_{l'}' \rangle&=\notag\\
\langle r_1| \ldots \langle r_l|\: ^{\text{\tiny 3}}\langle x_3|\:^{\text{\tiny 2}}
\langle x_2|\:^{\text{\tiny 1}}\langle x_1|x_1'\rangle^{\text{\tiny 1}}\:|x_2'
\rangle^{\text{\tiny 2}}\:|x_3'\rangle^{\text{\tiny 3}}\:|r_1'\rangle \ldots |r_{l'}' 
\rangle&=\nn\\
\delta(x_1-x_1')\; \delta(x_2-x_2')\; \delta_{l,l'}
\sum_{\pi \in S_l}\prod_{k=1}^{l}& \delta(r_{\pi(k)}-r_k') \;r_k'
\end{align}
Inner products between two different classes of states 
(${\cal O}_{3;l},{\cal O}_{2,1,;l},{\cal O}_{1,1,1;l}$) vanish.

In the strict planar limit $(g_2=0$)
$H$ can be regarded as a free quantum mechanical system, with the
following action on the continuum states
\begin{align}
&{\it {H}}_0 \, |x_1, x_2, r_0-x_1-x_2\rangle^{\text{\tiny 123}}|r_1\rangle 
\ldots |r_l \rangle \notag\\ 
&\qquad
=-\frac{\lambda'}{8 \pi^2} 
\left[ \left(\p_{x_1} - \p_{x_2}\right)^2 + \p_{x_1}^2 + \p_{x_2}^2\right]
|x_1, x_2, r_0-x_1-x_2\rangle^{\text{\tiny 123}}|r_1\rangle \ldots |r_l \rangle\nn\\&
\end{align}
and
\[
{\it {H}}_0 \, |x,r_0-s-x\rangle^{\text{\tiny ab}}|s\rangle^{\text{\tiny c}}
|r_1\rangle \ldots |r_l \rangle 
=-\frac{\lambda'}{4 \pi^2}\; \p_x^2\;|x,r_0-s-x\rangle^{\text{\tiny ab}}|s\rangle^{\text{\tiny c}}|r_1\rangle \ldots |r_l \rangle\, ,
\]
This last expression is equivalent to the action of $H_0$ on two-impurity states 
\cite{QM} -- the third impurity $|s\rangle^c$ just gets carried along. 
Note that the ${\cal O}_{1,1,1;l}$ state 
$|x_1\rangle^{\text{\tiny 1}}
|x_2\rangle^{\text{\tiny 2}}|x_3\rangle^{\text{\tiny 3}}
|r_1\rangle \ldots |r_l\rangle$
is annihilated by $H_0$ as stated in \eqref{h_o(2,1)}.

This eigenvalue problem is solved by the following
momentum states \cite{Boston7,planar3imp}
\begin{align} \label{eigenstates1}
&|n_1, n_2; r_0 \rangle |r_1 \rangle...|r_l \rangle =
\frac{1}{r_0 \sqrt{r_1 \ldots r_l}}\notag \\
&\qquad\times
 \int_{0 \le x_1,x_2}^{x1+x2\le r_0} dx_1 dx_2
\Big( e^{2 \pi i (n_1 x_1+n_2(x_1+x_2))/r_0}|x_1, x_2, r_0-x_1-x_2
\rangle^{\text{\tiny 123}} \notag \\
&\qquad\qquad\qquad
+ e^{2 \pi i (n_2 x_1+n_1(x_1+x_2))/r_0}|x_1, x_2, r_0-x_1-x_2
\rangle^{\text{\tiny 132}}\Big)|r_1 \rangle... |r_l \rangle
\end{align}
and
\begin{align} \label{eigenstates2}
&|n; r_0-s \rangle^{\text{\tiny ab}}|s\rangle^{\text{\tiny c}} 
|r_1 \rangle... |r_l \rangle 
=\frac{1}{\sqrt{(r_0-s)r_1 \ldots r_l}}\nn\\
&\qquad\qquad\times \int_{0}^{r_0-s} dx\,
e^{2 \pi i n x/(r_0-s)}|x,r_0-x-s\rangle^{\text{\tiny ab}}|s\rangle^{\text{\tiny c}} 
|r_1 \rangle... |r_l \rangle \, .
\end{align}
They have the energy eigenvalues
\begin{align}
&E_{|n_1, n_2; r_0 \rangle|r_1 \rangle \ldots |r_l \rangle}=
\frac{\l'}{2 \, r_0^2}\,(n_1^2 + n_2^2 + (n_1+n_2)^2)
=\frac{\l'}{r_0^2}(n_1^2+n_2^2 +n_1\,n_2)\label{EnEV}\\
&E_{|n; r_0-s \rangle^{\text{\tiny ab}}|s\rangle^{\text{\tiny c}} 
|r_1 \rangle \ldots |r_l \rangle}=\frac{\l'\,n^2}{(r_0-s)^2}\;.
\end{align}
Trivially $|x_1\rangle^{\text{\tiny 1}}
|x_2\rangle^{\text{\tiny 2}}|x_3 \rangle^{\text{\tiny 3}}$ is 
an eigenstate of $H_0$ with zero eigenvalue. 
The scalar product of the above eigenstates follows from \eqref{ips}
\begin{align}
&\langle r_1| \ldots \langle r_l|\; \langle n_1,n_2;r_0|m_1,m_2;r_0'\rangle\;
|r_1'\rangle \ldots |r_{l'}' \rangle = 
\delta_{n_1,m_1}\; \delta_{n_2,m_2}\; \delta_{l,l'}\; (\Delta')^l_{r,r'}\nn\\
&\langle r_1| \ldots \langle r_l|\: ^{\text{\tiny c}}\langle s|\: ^{\text{\tiny ab}}
\langle n; r_0 - s |m; r_0'-s'\rangle^{\text{\tiny ab}}\:|s'\rangle^{\text{\tiny c}}
\:|r_1'\rangle \ldots |r_{l'}' \rangle =\nn\\
&\qquad\qquad\qquad\qquad\qquad\qquad\qquad\qquad\qquad\qquad
\delta_{n,m}\; \delta(s-s')\; \delta_{l,l'}\;(\Delta')^l_{r,r'}\;,
\end{align}
where $(\Delta')^l_{r,r'}=\sum_{\pi \in S_l}\prod_{k=1}^{l} 
\delta(r_{\pi(k)}-r_k')$.\footnote{It should be 
mentioned, that if only one of the two contributions 
in \eqref{eigenstates1} is included, these eigenstates are not orthogonal 
due to the condition $0 \le x_1+x_2 \le r_0$. 
This can be seen for finite $J$ as well: The corresponding BMN
operator is obtained by replacing three $Z$'s in the trace ${\rm Tr}\;
Z^{J+3}$ by three scalar impurities in all possible ways with
appropriate prefactors. The first part of \eqref{eigenstates1},
proportional to \eqref{o(3)}, represents only half of the
possibilities, because it does not contain the cases where two
impurities are exchanged.}

It is straightforward, yet rather tedious, to calculate the
action of $H_+$ on the $H_0$ eigenstates from the discrete system, 
one finds
\begin{align}  
&{\it {H}}_+ \,|n_1, n_2; r_0\rangle|r_1\rangle \ldots |r_l \rangle = 
\; \frac{\l'\, g_2}{2 \pi^3}\bigg(-\int_0^{r_0}dr_{l+1} \sum_{m_1,m_2}
\notag \\
&  \bigg(
\frac{\sqrt{r_{l+1}}}{r_0-r_{l+1}}\sin (\pi n_1 \tfrac{r_0-r_{l+1}}{r_0})
\sin (\pi n_2 \tfrac{r_0-r_{l+1}}{r_0})\sin (\pi (n_1+n_2) 
\tfrac{r_0-r_{l+1}}{r_0})\notag \\
&\times
\bigg[\frac{m_1(-m_1-2 m_2)}{m_1-n_1\tfrac{r_0-r_{l+1}}{r_0}}+
\frac{m_2(2m_1+ m_2)}{m_2-n_2\tfrac{r_0-r_{l+1}}{r_0}}+
\frac{-(m_1+m_2)(m_2-m_1)}{-(m_1+m_2)+(n_1+n_2)\tfrac{r_0-r_{l+1}}{r_0}}\bigg]\notag\\
&\quad\quad\quad\quad\quad\quad\quad\quad \times
\frac{1}{m_1 n_2-m_2 n_1}\bigg)|m_1, m_2; r_0-r_{l+1}\rangle |r_1\rangle 
\ldots |r_{l+1}\rangle\notag\\
&
+ \int_0^{r_0}ds\; \sum_m  \frac{m}{\sqrt{r_0-s}} \sin (\pi n_1 \tfrac{s}{r_0})
\sin (\pi n_2 \tfrac{s}{r_0})\sin (\pi (n_1+n_2) \tfrac{s}{r_0})\notag \\
&\quad
 \times \sum_{(a,b)(c)} \frac{1}{N_c}\Big(\frac{1}{m + N_a \tfrac{r_0-s}{r_0}}+
\frac{1}{m - N_b \tfrac{r_0-s}{r_0}}\Big)
 |m; r_0-s\rangle^{\text{\tiny ab}}|s\rangle^{\text{\tiny c}}|r_1\rangle \ldots 
|r_l \rangle\bigg)\nn\\
\end{align}
where the sum over $(a,b)(c)$ runs over the triples $\{(1,2)(3);(2,3)(1);(1,3)(2)\}$ 
as in \eqref{o(2,1)}. Moreover we have
\begin{align}
&{\it {H}}_+|n;r_0-s\rangle^{\text{\tiny ab}}|s\rangle^{\text{\tiny c}}
|r_1\rangle \ldots |r_l \rangle=
\frac{\l'\, g_2}{\pi^2} \int_0^{r_0-s}dr_{l+1}\sum_m 
\sqrt{\frac{r_{l+1}}{(r_0-s-r_{l+1})(r_0-s)}}\,
\notag\\
&\qquad\qquad
\times 
\frac{m \sin^2(\pi n \frac{r_0-s-r_{l+1}}{r_0-s})}{m-n\tfrac{r_0-s-r_{l+1}}{r_0-s}}|m;r_0-s-r_{l+1}\rangle^{\text{\tiny ab}}|s\rangle^{\text{\tiny c}}|r_1\rangle .. |r_{l+1} \rangle
\end{align}   
where $t_1+t_2+t_3=r_0$ and $N_1:=-(n_1+n_2)$,\;$N_2:=n_1$,\;$N_3:=n_2$.
The action of $H_-$ on our eigenstates is stated in Appendix \ref{H&S}.


\subsection{Matrix elements of $\S$} 

As discussed in the introduction the transition to the string Hamiltonian is performed
with the help of the Hermitian operator $S$, which induces the complete tree-level
two point functions. We shall need only the linear term in a small $g_2$
expansion, $S=\mathbf{1}+g_2\,\Sigma+{\cal O}(g_2{}^2)$. It has been conjectured 
in \cite{Verlinde} that $S$ exponentiates, i.e. $S=\exp[g_2\, \Sigma]$.

The matrix elements of $\Sigma$ in the momentum basis can be computed from the
free planar two point functions of $k$ and $k+1$ trace operators, i.e.
one needs to know the correlators
\begin{align}
&\langle {\mathcal O}_{3;l}\bar{{\mathcal O}}_{3;l+1}\rangle_{0}\,,\;
\langle {\mathcal O}_{3;l}\bar{{\mathcal O}}_{2,1;l}\rangle_{0}\,, \;
\langle {\mathcal O}_{2,1;l}\bar{{\mathcal O}}_{2,1;l+1}\rangle_{0}\nn\\
&\langle {\mathcal O}_{2,1;l}\bar{{\mathcal O}}_{1,1,1;l}\rangle_{0}\,,\; 
\langle {\mathcal O}_{1,1,1;l}\bar{{\mathcal O}}_{1,1,1;l+1}\rangle_{0}
\end{align}          
at leading order in $N$ and then take the $J\to\infty$ limit. 

The operator $\Sigma$ again splits into a trace number increasing ($\Sigma_+$) and
decreasing ($\Sigma_-$) piece. Acting with $\Sigma_+$ on our three classes of momentum eigenstates 
we find 
\begin{align}
&\Sigma_+ |n_1,n_2;r_0\rangle |r_1\rangle \ldots |r_l\rangle
=\int_0^{r_0}dr_{l+1} \sum_{m_1,m_2}
\notag\\
&\qquad
\times\frac{\sin(\pi n_1\tfrac{r_0-r_{l+1}}{r_0})\sin(\pi n_2
\tfrac{r_0-r_{l+1}}{r_0})\sin(\pi\left(n_1+n_2\right)\tfrac{r_0-r_{l+1}}{r_0})}
{(m_1-n_1\tfrac{r_0-r_{l+1}}{r_0})(m_2-n_2\tfrac{r_0-r_{l+1}}{r_0})(-(m_1+m_2)+(n_1+n_2)
\tfrac{r_0-r_{l+1}}{r_0})}\notag\\
&\qquad\qquad\qquad\times
\frac{\left(r_0-r_{l+1}\right)^2\sqrt r_{l+1}}{r_0 \pi^3}\;|m_1,m_2;r_0-r_{l+1}
\rangle |r_1\rangle \ldots |r_{l+1}\rangle\notag\\
&\; +
\frac{1}{2} \sum_{i=1}^l \int_0^{r_i}dr_{l+1} \sqrt{r_i(r_i-r_{l+1})r_{l+1}}\;|n_1,n_2;r_0
\rangle |r_1\rangle \ldots |r_i-r_{l+1}\rangle \ldots |r_{l+1}\rangle\notag\\
&\; +
\int_0^{r_0}ds \sum_m \frac{(r_0-s)^{3/2}}{\pi^3} \sin(\pi n_1 \tfrac{s}{r_0}) \sin(\pi n_2 
\tfrac{s}{r_0}) \sin(\pi \left(n_1+n_2\right) \tfrac{s}{r_0})\notag\\
&\quad\times
\sum_{(a,b)(c)} \left(\left(m+N_a\tfrac{r_0-s}{r_0}\right)\left(m-N_b\tfrac{r_0-s}{r_0}\right)N_c
\right)^{-1}\;|m;r_0-s\rangle^{\text{\tiny ab}}|s\rangle^{\text{\tiny c}}|r_1\rangle \ldots |r_{l}\rangle
\end{align}
where again the sum $(ab)(c)$ runs over the triples $\{(1,2)(3);(2,3)(1);(1,3)(2)\}$ and
\begin{align} \label{S1}
&\Sigma_+
|n;r_0-s\rangle^{\text{\tiny ab}}|s\rangle^{\text{\tiny c}}|r_1\rangle \ldots |r_{l}\rangle
=\int_0^{r_0-s}dr_{l+1} 
\sum_m\notag\\
&\, \sqrt{ \frac{r_{l+1}(r_0-s-r_{l+1})^3}{r_0-s}}\frac{\sin^2(\pi n 
\tfrac{r_0-s-r_{l+1}}{r_0-s})}{\pi^2(m-n \tfrac{r_0-s-r_{l+1}}{r_0-s})^2}\,|m;r_0-s-r_{l+1}
\rangle^{\text{\tiny ab}}|s\rangle^{\text{\tiny c}}|r_1\rangle...|r_{l+1}\rangle\notag \\
&\; + 
\int_0^s dr_{l+1}\sqrt{r_{l+1}}\;(s-r_{l+1})\;|n;r_0-s\rangle^{\text{\tiny ab}}|s-r_{l+1}
\rangle^{\text{\tiny c}}|r_1\rangle \ldots |r_{l+1}\rangle\notag\\ 
&\; +
\frac{1}{2} \sum_{i=1}^l \int_0^{r_i}dr_{l+1} \sqrt{r_i(r_i-r_{l+1})r_{l+1}}
\;|n;r_0-s\rangle^{\text{\tiny ab}}|s\rangle^{\text{\tiny c}}|r_1\rangle...|r_i-r_{l+1}
\rangle...|r_{l+1}\rangle\notag\\
&\; -
\int_0^{r_0-s} dt \frac{(r_0-s)^{3/2}}{n^2\; \pi^2} \sin^2 (\pi n \tfrac{t}{r_0-s})\;|t
\rangle^{\text{\tiny a}}|r_0-s-t\rangle^{\text{\tiny b}}|s\rangle^{\text{\tiny c}}|r_1
\rangle \ldots |r_{l}\rangle 
\end{align}
as well as
\begin{align}\label{S2}
&\Sigma_+|t_1\rangle^{\text{\tiny 1}}|t_2\rangle^{\text{\tiny 2}}|t_3\rangle^{\text{\tiny 3}}|r_1\rangle \ldots |r_{l}\rangle=\notag\\
&\; \int_0^{t_1} dr_{l+1} \left(t_1-r_{l+1}\right)\sqrt{r_{l+1}}\;|t_1-r_{l+1}\rangle^{\text{\tiny 1}}|t_2\rangle^{\text{\tiny 2}}|t_3\rangle^{\text{\tiny 3}}|r_1\rangle \ldots |r_{l+1}\rangle\notag\\
&\quad +
\int_0^{t_2}dr_{l+1} \left(t_2-r_{l+1}\right)\sqrt{r_{l+1}}\;|t_1\rangle^{\text{\tiny 1}}|t_2-r_{l+1}\rangle^{\text{\tiny 2}}|t_3\rangle^{\text{\tiny 3}}|r_1\rangle \ldots |r_{l+1}\rangle\notag\\
&\quad + 
\int_0^{t_3}dr_{l+1} \left(t_3-r_{l+1}\right)\sqrt{r_{l+1}}\;|t_1\rangle^{\text{\tiny 1}}|t_2\rangle^{\text{\tiny 2}}|t_3-r_{l+1}\rangle^{\text{\tiny 3}}|r_1\rangle \ldots |r_{l+1}\rangle\notag\\
&\quad +
\sum_{i=1}^l \int_0^{r_i} dr_{l+1}\sqrt{r_i(r_i-r_{l+1})r_{l+1}}\;|t_1\rangle^{\text{\tiny 1}}|t_2\rangle^{\text{\tiny 2}}|t_3\rangle^{\text{\tiny 3}}|r_1\rangle \ldots |r_i-r_{l+1}\rangle \ldots |r_{l+1}\rangle
\end{align}
where $t_1+t_2+t_3=r_0$ and $N_1:=-(n_1+n_2)$,\;$N_2:=n_1$,\;$N_3:=n_2$. Note that \eqref{S1} and 
\eqref{S2} are again very similar to the corresponding expressions for two impurities computed
in \cite{SV}.
The action of $\S_-$ on these states, which also follow form Hermiticity, can be found in 
Appendix \ref{H&S}.

In the case of two impurities it turned out to be very useful to work with an operator
$Q_0$ being the square root of $H_0$, as then the remarkable relation  
$
H_{\pm} = Q_0 [Q_0,\S_{\pm}]\,
$
could be proved \cite{SV}. Here, in the case of three impurities, an analogue relation
does not hold, essentially because now the energy eigenvalues \eqref{EnEV} are not perfect squares.


\subsection{The String Hamiltonian $\tilde{H}$}

We have now assembled all the necessary ingredients to establish the form of
the string Hamiltonian at order $g_2$
\[
\langle a |\tilde{H}|b\rangle  
=\langle a| H_0|b\rangle  + g_2\, \langle a| \tfrac{1}{2}\, [\S,H_0] + H_- + H_+| b \rangle\,.    
\label{Htildedef}
\]
Explicitly we find for the action of $\tilde{H}$ on the eigenstates
\begin{align} \label{h_tilde_matrixelement}
&{\it \tilde{H}}|n_1,n_2;r_0\rangle |r_1\rangle \ldots |r_l\rangle =\notag\\
&\;\frac{\lambda'}{r_0^2}(n_1^2+n_2^2+n_1 n_2)|n_1,n_2;r_0\rangle |r_1\rangle \ldots |r_l\rangle\notag\\
&+\frac{\lambda'g_2}{2 \pi^3} \bigg[
\int_0^{r_0} \hspace{-1mm} d r_{l+1} \hspace{-1mm} \sum_{m_1,m_2} \hspace{-1mm}
\Big[\sin (\pi n_1 \tfrac{r_0-r_{l+1}}{r_0}) \sin (\pi n_2 \tfrac{r_0-r_{l+1}}{r_0})
\sin (\pi (n_1+n_2) \tfrac{r_0-r_{l+1}}{r_0})\notag\\
&\quad
\times\Big(\frac{1}{m_1-n_1\tfrac{r_0-r_{l+1}}{r_0}}+\frac{1}{m_2-n_2\tfrac{r_0-r_{l+1}}{r_0}}+
\frac{1}{-(m_1+m_2)+(n_1+n_2)\tfrac{r_0-r_{l+1}}{r_0}}\Big)\notag\\
&\quad
\times\Big(-\frac{\sqrt{r_{l+1}}}{r_0}\Big)  |m_1, m_2; r_0-r_{l+1}\rangle |r_1\rangle 
\ldots |r_{l+1}\rangle\Big]\notag\\ 
&\;
+\sum_{i=1}^l\sum_{m_1,m_2}\Big[
\sin (\pi m_1 \tfrac{r_0}{r_0+r_i})\sin (\pi m_2 \tfrac{r_0}{r_0+r_i})\sin (\pi (m_1+m_2) 
\tfrac{r_0}{r_0+r_i}) \notag\\
&\quad
\times\Big(\frac{1}{m_1-n_1\tfrac{r_0+r_{i}}{r_0}}+\frac{1}{m_2-n_2\tfrac{r_0+r_{i}}{r_0}}+
\frac{1}{-(m_1+m_2)+(n_1+n_2)\tfrac{r_0+r_{i}}{r_0}}\Big)\notag\\
&\quad
\times\Big(\frac{\sqrt{r_{i}}}{r_0}\Big)|m_1, m_2; r_0+r_{i}\rangle |r_1\rangle 
\ldots\makebox[0pt]{\;\;$\not$}|r_i\rangle \ldots |r_{l}\rangle\Big]\notag\\
&\;
+\int_0^{r_0} ds \sum_m \frac{\sqrt{r_0-s}}{r_0}\sin (\pi n_1 \tfrac{r_0-s}{r_0}) 
\sin (\pi n_2 \tfrac{r_0-s}{r_0})\sin (\pi (n_1+n_2) \tfrac{r_0-s}{r_0})\notag\\
&\quad
\times \hspace{-1mm} \sum_{(a,b)(c)} \hspace{-1mm}
\Big(\frac{1}{N_a \tfrac{r_0-s}{r_0}+m}+\frac{1}{N_b \tfrac{r_0-s}{r_0}-m}+
\frac{1}{N_c \tfrac{r_0-s}{r_0}}\Big)|m;r_0-s\rangle^{\text{\tiny ab}}|s\rangle^{\text{\tiny c}}
|r_1\rangle ... |r_l \rangle\bigg]
\end{align}
as well as
\begin{align}
&{\it \tilde{H}}|n;r_0-s\rangle^{\text{\tiny ab}}|s\rangle^{\text{\tiny c}}
|r_1\rangle \ldots |r_l \rangle=\notag\\
&\;\frac{\lambda'}{(r_0-s)^2}n^2|n;r_0-s\rangle^{\text{\tiny ab}}|s\rangle^{\text{\tiny c}}
|r_1\rangle \ldots |r_l \rangle\notag\\
&+ \frac{\lambda'g_2}{2 \pi^2} \bigg[\int_0^{r_0-s}dr_{l+1}\sum_m 
\sqrt{\frac{r_{l+1}}{(r_0-s-r_{l+1})(r_0-s)}}\notag\\
&\qquad\; \times\sin^2(\pi n \frac{r_0-s-r_{l+1}}{r_0-s})
|m;r_0-s-r_{l+1}\rangle^{\text{\tiny ab}}|s\rangle^{\text{\tiny c}}|r_1\rangle .. |r_{l+1} \rangle\notag\\
&\quad + \sum_{i=1}^l \sum_m\sqrt{\frac{r_i}{(r_0-s+r_i)(r_0-s)}}\nn\\
&\qquad\;  \times\sin^2(\pi m \frac{r_0-s}{r_0-s+r_i})
|m;r_0-s+r_i\rangle^{\text{\tiny ab}}|s\rangle^{\text{\tiny c}}
|r_1\rangle..\makebox[0pt]{\;\;$\not$}|r_i\rangle..|r_l\rangle\notag\\
&\quad + \sum_{m_1,m_2} \frac{\sqrt{r_0-s}}{ \pi\; r_0}\sin (\pi m_1 \tfrac{r_0-s}{r_0}) 
\sin (\pi m_2 \tfrac{r_0-s}{r_0})\sin (\pi (m_1+m_2) \tfrac{r_0-s}{r_0})\notag\\
&\qquad\;\times\hspace{-1mm} \sum_{(a,b)(c)}\hspace{-1mm}\Big(\frac{1}{M_a \tfrac{r_0-s}{r_0}+n}
+ \frac{1}{M_b \tfrac{r_0-s}{r_0}-n} 
+ \frac{1}{M_c \tfrac{r_0-s}{r_0}}\Big)|m_1,m_2;r_0\rangle|r_1\rangle \ldots |r_l \rangle\notag\\
&\quad - \int_0^{r_0-s} dt \frac{1}{\sqrt{r_0-s}} \sin^2 (\pi n \tfrac{t}{r_0-s})
|t\rangle^{\text{\tiny a}}|r_0-s-t\rangle^{\text{\tiny b}}|s\rangle^{\text{\tiny c}}
|r_1\rangle \ldots |r_{l}\rangle\bigg]
\end{align}
and finally
\begin{align}
&{\it \tilde{H}}|t_a\rangle^{\text{\tiny a}}|t_b\rangle^{\text{\tiny b}}|t_c\rangle^{\text{\tiny c}}
|r_1\rangle \ldots |r_{l}\rangle =\notag\\
&\;-\frac{\lambda' g_2}{2 \pi^2}\sum_m\sum_{(a,b)(c)}\frac{1}{\sqrt{t_a+t_b}}
\sin^2(\pi m \tfrac{t_a}{t_a+t_b})|m;t_a+t_b\rangle^{\text{\tiny ab}}|t_c\rangle^{\text{\tiny c}}
|r_1\rangle \ldots |r_{l}\rangle\;,
\end{align}
where we have used the same definitions as in the previous sections. One may easily check that
$\tilde H$ is a Hermitian operator. In the following chapter
we will check that these matrix elements agree with string field theory calculations.  


\section{Comparison with Light-Cone String Field Theory}
\setcounter{equation}{0}

Light-cone string field theory for superstrings in the maximally
supersymmetric plane-wave background has been
developed in a number of works \cite{SFT}. For our purposes we shall make use of a compact expression
for the three-string interaction vertex involving only bosonic excitations, as
derived  in \cite{Bonderson}. The interaction matrix element in question
is described by a three-string state made of $2k$ purely 
bosonic excitation operators
\[
|A\rangle = \prod_{j=1}^{2k}\, \alpha_{(r_j)\, m_j}^{I_j\, \dagger}\, |0\rangle
\]
with (say) the transverse space index $I_j\in(1,\ldots, 4)$, 
$r_j\in(1,2,3)$ denoting the string number and $m_j$ the
associated excitation modes. We wish to compare our results
of \eqref{h_tilde_matrixelement}
to the $3\rightarrow 2+1$ and the $3\rightarrow 3$ transition
amplitudes.

In the first case we are dealing with a three-string state of the form
\[
|\psi_{3\rightarrow 2+1}\rangle = \alpha^{1\, \dagger}_{(1)\, -m}\,
\alpha^{2\, \dagger}_{(1)\, m}\, \alpha^{3\, \dagger}_{(2)\, 0}\,
\alpha^{1\,\dagger}_{(3)\, -n_1-n_2}\, \alpha^{2\,\dagger}_{(3)\, n_1}\,
\alpha^{3\, \dagger}_{(3)\, n_2}\, |0\rangle\, ,
\]
making use of the vertex presented in equation (15) of \cite{Bonderson} one finds
for its matrix element
\begin{align}
\langle \psi_{3\rightarrow 2+1} |H_3\rangle  =\ft{\alpha_{(1)}\alpha_{(3)}\alpha_{(3)}}{2}\,
\tilde N^{2,3}_{0,n_2}\, \Bigl [ &\left ( 
\frac{\omega_{(1)\, m}}{\mu\, \alpha_{(1)}} + 
\frac{\omega_{(3)\, -n_1-n_2}}{\mu\, \alpha_{(3)}} \right )
\, \tilde N^{1,3}_{-m,-n_1-n_2}\, \tilde N^{1,3}_{-m,n_1} \nn\\
+ &\left (
\frac{\omega_{(1)\, m}}{\mu\, \alpha_{(1)}} + \frac{\omega_{(3)\, n1}}{\mu\, \alpha_{(3)}} \right )
\, \tilde N^{1,3}_{m,-n_1-n_2}\, \tilde N^{1,3}_{m,n_1} \nn\\ +&
\left (
\frac{\omega_{(2)\, 0}}{\mu\, \alpha_{(2)}} + \frac{\omega_{(3)\, n2}}{\mu\, \alpha_{(3)}} \right )
\, \tilde N^{1,3}_{-m,-n_1-n_2}\, \tilde N^{1,3}_{-m,n_1} \Bigr ] \, ,
\label{3->2+1}
\end{align}
where the $r$'th string
frequencies are given by $\omega_{(r)\, m}=\sqrt{m^2+\mu^2\, \alpha_{(r)}^2}$
and the fractions of $p^+$ momenta for the three-strings  read
$\alpha_{(1)}=(1-s)$, $\alpha_{(2)}=s$ and $\alpha_{(3)}=-1$
in our conventions. Moreover the Neumann
matrices $\tilde N^{r,s}_{m,n}$ are given by ($m,n\neq0$) \cite{HSSV}\footnote{There
are corrections to these formulas of the form $\exp(-2\pi|\alpha_{(r)}|\mu)$, which have
been computed recently \cite{sakura}. These, however, are not effective in the large $\mu$
limit we are considering.} 
\begin{align}
\tilde N^{r,s}_{0,n} &= \ft 1{\sqrt{2}}\, \bar N^{r,s}_{0,|n|} \nn\\
\tilde N^{r,s}_{m,n} &= \ft 1 2\,\left ( \,\bar N^{r,s}_{|m|,|n|}
-\mbox{sign} (m\cdot n)\, \bar N^{r,s}_{-|m|,-|n|}\, \right )\nn\\
\bar N^{r,s}_{0,|n|}&= \frac{(-1)^{s(|n|+1)}\, s_{(s)\, n}}
{2\pi}
\sqrt{\frac{|\alpha_{(s)}|}
{\alpha_{(r)}\, \omega_{(s)\,n} (\omega_{(s)n}+\mu\, \alpha_{(s)})}}\nn\\
\bar N^{r,s}_{\pm|m|,\pm|n|}&= \pm\frac{(-1)^{r(|m|+1)+s(|n|+1)}s_{(r)}\, s_{(s)\, n}}
{2\pi(\alpha_{(s)}\omega_{(r)\, m}
-\alpha_{(r)}\omega_{(s)\, n})}\, \nn\\&\qquad \times
\sqrt{\frac{|\alpha_{(r)}\, \alpha_{(s)}|\,
(\omega_{(r)m}\pm\mu\, \alpha_{(r)})\, (\omega_{(s)n}\pm\mu\, \alpha_{(s)})}{\omega_{(r)\, m}\,
\omega_{(s)\, n}}}\, \nn\\
s_{(1)\,m}&= 1= s_{(2)\, m}\qquad s_{(3)\, m} =  2\sin\left (\pi\, |m|\, 
\frac{\alpha_{(1)}}{\alpha_{(3)}} \right )
\end{align}
Expanding out \eqref{3->2+1} to leading order in $\ft{1}{\mu}$ using the above formulas
yields
\begin{align}
\langle \psi_{3\rightarrow 2+1} |H_3\rangle  &=
-\frac{\sqrt{s}\, (1-s)}{2\pi^3\, \mu^2}\, \sin(\pi\, n_1\, s)\, \sin(\pi\, n_2\, s)\, 
\sin(\pi\, (n_1+n_2)\, s)\, \nn\\
&\times 
\left [ \frac{1}{(-n_1-n_2)(1-s)+m}+\frac{1}{n_1(1-s)-m}
+\frac{1}{n_2(1-s)} \right ] \nn\\
&\qquad + {\cal O}(\frac{1}{\mu^4})
\end{align}
which is (up to a normalization factor of $\sqrt{s(1-s)}$\, ) precisely equal
to the gauge theory result of the last two lines of \eqref{h_tilde_matrixelement}!

Similarly one obtains for the $3\rightarrow 3$ matrix element associated to the
three-string state
\[
|\psi_{3\rightarrow 3}\rangle = \alpha^{1\, \dagger}_{(1)\, m_1}\,
\alpha^{2\, \dagger}_{(1)\, m_2}\, \alpha^{3\, \dagger}_{(1)\, -m_1-m_2}\,
\alpha^{1\,\dagger}_{(3)\, n_1}\, \alpha^{2\,\dagger}_{(3)\, n_2}\,
\alpha^{3\, \dagger}_{(3)\, -n_1-n_2}\, |0\rangle
\]
the amplitude
\begin{align}
\langle \psi_{3\rightarrow 3} |H_3\rangle  &=\ft{\alpha_{(1)}\alpha_{(3)}\alpha_{(3)}}{2}
\times\,\nn\\
\Bigl [ &\left ( 
\frac{\omega_{(1)\, m_1}}{\mu\, \alpha_{(1)}} + 
\frac{\omega_{(3)\, n_1}}{\mu\, \alpha_{(3)}} \right )
\, \tilde N^{1,3}_{-m_1,n_1}\, \tilde N^{1,3}_{m_2,n_2} \, \tilde N^{1,3}_{-m_1-m_2,-n_1-n_2} \nn\\
+ &
\left ( 
\frac{\omega_{(1)\, m_2}}{\mu\, \alpha_{(1)}} + 
\frac{\omega_{(3)\, n_2}}{\mu\, \alpha_{(3)}} \right )
\, \tilde N^{1,3}_{m_1,n_1}\, \tilde N^{1,3}_{-m_2,n_2} \, \tilde N^{1,3}_{-m_1-m_2,-n_1-n_2} \nn\\
+ &
\left ( 
\frac{\omega_{(1)\, m_1+m_2}}{\mu\, \alpha_{(1)}} + 
\frac{\omega_{(3)\, n_1+n_2}}{\mu\, \alpha_{(3)}} \right )
\, \tilde N^{1,3}_{m_1,n_1}\, \tilde N^{1,3}_{m_2,n_2} \, \tilde N^{1,3}_{m_1+m_2,-n_1-n_2} 
\, \Bigr ]\, ,
\label{3->3}
\end{align}
which in the $\mu\to\infty$ limit reduces to
\begin{align}
\langle \psi_{3\rightarrow 3} |H_3\rangle  &=
\frac{s \sqrt{1-s}}{2\pi^3\, \mu^2}\, \sin(\pi\, n_1\, s)\, \sin(\pi\, n_2\, s)\, 
\sin(\pi\, (n_1+n_2)\, s)\, \nn\\
&\times 
\Bigl [ \frac{1}{m_1-n_1(1-s)}+\frac{1}{m_2-n_2(1-s)}\nn\\ &\qquad 
+ \frac{1}{-(m_1+m_2)+(n_1+n_2)(1-s)} \Bigr ] + {\cal O}(\frac{1}{\mu^4}) \, .
\end{align}
This result similarly agrees with our gauge theory findings in \eqref{h_tilde_matrixelement}
modulo the identical normalization factor of $\sqrt{s(1-s)}$. This concludes our investigations
on the dual string field theory side.


\section{Decay of a single trace state}
\setcounter{equation}{0}

Finally we turn to the evaluation of the decay widths of a single trace (string)
state into the continuum of degenerate 
double trace (string) states. A given state $|n_1,n_2;1\rangle$ 
has two possible decay channels, namely 
\begin{enumerate}
 \item\quad $|n_1,n_2;1\rangle \rightarrow |m_1,m_2;1-r\rangle|r\rangle$ 
 \item\quad $|n_1,n_2;1\rangle \rightarrow |m;1-s\rangle^{\text{\tiny ab}}
|s\rangle^{\text{\tiny c}}$
\end{enumerate}
where in both cases the final state spectrum is continuous.
The decay width for each channel can be computed in terms of quantum mechanical 
time-dependent perturbation theory. At leading order the decay width is given by
\begin{align} \label{decaywidth}
\G = \sum_f 2 \pi \big|\langle f|\tilde{H}|i\rangle\big|^2 \d(E_i-E_f)
\end{align}
where $\langle f|\tilde{H}|i\rangle$ is the transition amplitude between the
initial ($i$) and a particular final ($f$) state and the sum $\sum_f$
runs over all possible final states, which are degenerate in
energy $E_i=E_f$.

We will
first concentrate on decay channel 1. There \eqref{decaywidth} reads
\begin{align} \label{g_(3;1)}
&\G^{(1)} = 2 \pi \sum_{m_1,m_2}\int_0^1 dr \;\big|\langle r|\langle m_1,m_2;1-r|
\tilde{H}|n_1,n_2;1\rangle\big|^2\;\nn\\&\qquad\qquad\qquad\qquad \times
 \d( E_{|n_1,n_2;1\rangle}-E_{|m_1,m_2;1-r\rangle|r\rangle})\notag\\\ \\
&\qquad\qquad\qquad\quad \text{with} \quad E_{|n_1,n_2;1\rangle} = 
\l'(n_1^2 + n_2^2 + n_1 n_2)  \notag\\
&\qquad\qquad\qquad\quad\, \text{and}  \quad E_{|m_1,m_2;1-r\rangle|r\rangle} = \tfrac{\l'}{(1-r)^2}(m_1^2 + m_2^2 + m_1 m_2)\nn
\end{align}
The $\d$-function can be rewritten into a $\d$-function for $r$
\begin{align} \label{d-function}
&\d( E_{|n_1,n_2;1\rangle}-E_{|m_1,m_2;1-r\rangle|r\rangle}) \longrightarrow \notag \\
&\qquad 
\frac{1}{2 \l'}\;\Big[\frac{m_1^2 + m_2^2 + m_1 m_2}{(n_1^2 + n_2^2 + n_1 n_2)^3}
\Big]^{1/2}\; \d\Big(r - \Big(1 - \Big[\frac{m_1^2 + m_2^2 + m_1 m_2}{n_1^2 + n_2^2 
+ n_1 n_2}\Big]^{1/2}\Big)\Big) \\ \label{condition}
&\qquad\qquad\qquad\qquad\quad \text{with} \quad m_1^2 + m_2^2 + m_1 m_2 \le n_1^2 
+ n_2^2 + n_1 n_2.
\end{align}
We have omitted a term with an opposite sign in the $\d$-function,
because it does not satisfy $r \in [0,1]$. For the same reason the
condition \eqref{condition} has to be imposed on the sums over $m_1$
and $m_2$ in \eqref{g_(3;1)}. The meaning of this condition can be
illustrated by analyzing the discrete energy spectrum of the states
$|n_1,n_2;1\rangle$ more carefully.\\ Consider the equation
\begin{equation} \label{ellipse}
x^2 + y^2 + x y = R^2.
\end{equation}
By expressing $(x,y)$ via the rotated coordinates $(x',y')$ 
\begin{equation}
R^2 = \frac{1}{2} \big((x'-y')^2 + (x'+y')^2 + (x'-y')(x'+ y')\big) = \frac{1}{2}\big(3 \,x'^2 + y'^2\big)
\end{equation}
one can easily see, that \eqref{ellipse} defines an ellipse, which is
rotated by $\tfrac{\pi}{4}$ about the origin. Hence the points
$(n_1,n_2)$ of states with degenerate energies lie on an ellipsis. Two
outstanding classes of states, which form symmetry axes of the spectrum
are
\begin{itemize}
 \item $(n_1,n_2) \in \{(n,0)\;;\;(0,n)\;;\;(n,-n)\}$    with $E = \l' n^2$ 
 \item $(n_1,n_2) \in \{(n,n)\;;\;(n,-2n)\;;\;(-2n,n)\}$ with $E = 3 \l' n^2$,
\end{itemize} 
They will appear again in the calculation of the decay widths. For
symmetry reasons it is obvious that the full information of the
spectrum is already contained in the area bordered by the lines
$(n,n)$ and $(n,-n)$. The structure of the spectrum is schematically
depicted in figure 1.
\begin{figure} 
\begin{center}
\includegraphics[height=2in]{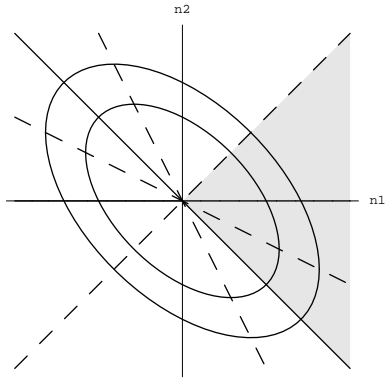}
\end{center}
\caption{Schematic structure of the energy spectrum. The ellipses 
correspond to levels of different energies, states with $E =\l' n^2$  
($E = 3 \l' n^2$) lie on continuous (dashed) lines. The shaded area 
contains the full information of the spectrum.}
\end{figure}
The condition \eqref{condition} can now be interpreted as the
restriction to the set of states, where $(m_1,m_2)$ lies within the
ellipse given by $E_{|n_1,n_2;1\rangle}$. Although this picture appears
to be rather simple, it remains a nontrivial problem to specify the
limits of the sums over $m_1$ and $m_2$ explicitly. However, this
problem can be solved at least numerically. The resulting
decay widths have been plotted in figure 2 for $n_1,n_2\in [-8,8]$. 
We will not describe this
method in detail and turn instead to the second decay channel, where
this problem does not occur.

For the decay channel 2 the condition
\eqref{condition} is
replaced by $|m| \le n_1^2 + n_2^2 + n_1 n_2$ and thus
\eqref{decaywidth} is given by
\begin{align} \label{G^2}
\G^{(2)} = 2 \pi \sum_{(a,b)(c)} \sum_{m=-[Q]}^{[Q]}\int_0^1 ds &
\;\big| \;^{\text{\tiny c}}\langle s|\;^{\text{\tiny ab}}\langle m;1-s|
\tilde{H}|n_1,n_2;1\rangle\big|^2\;\nn\\&\times
 \d( E_{|n_1,n_2;1\rangle}-E_{|m;1-s\rangle|s\rangle})\;,
\end{align}
where $Q \equiv \sqrt{n_1^2 + n_2^2 + n_1 n_2}$ and $[Q]$ denotes the
integer part of $Q$. Plugging in the matrix element of $\tilde{H}$ from
\eqref{h_tilde_matrixelement} and rewriting the $\d$-function in the
same manner as it was done in \eqref{d-function} leads to
\begin{align}
&\G^{(2)}=\frac{\l' \, g_2^2}{4 \pi^5}\frac{1}{Q^2}\, A_{n_1,n_2}\, B_{n_1,n_2} 
\qquad \text{with}\; \\
&
A_{n_1,n_2}= \sum_{m=1}^{[Q]}\, \sin^2 \big(\pi \, n_1 \, \frac{m}{Q} 
\, \big) \, \sin^2 \big(\pi \, n_2 \, \frac{m}{Q} \, \big) \, \sin^2 
\big(\pi \, (n_1+n_2)\, \frac{m}{Q} \, \big)\, ,\notag\\
&
B_{n_1,n_2} =\hspace{-1mm} \sum_{(a,b)(c)}\hspace{-1mm} 
\Big\{ \Big(\frac{1}{N_a + Q} + \frac{1}{N_b - Q} + \frac{1}{N_c} \Big)^2 +
\Big(\frac{1}{N_a - Q} + \frac{1}{N_b + Q} + \frac{1}{N_c} \Big)^2 \Big\} \notag\\
&\phantom{B_{n_1,n_2}}    = \, 18 \, \frac{Q^4}{n_1^2 \, n_2^2 
\, (n_1 + n_2)^2}\notag 
\end{align} 
after the integration over $s$. The sum $A_{n_1,n_2}$ is calculated in 
Appendix B. We finally find the result
\[\label{gamma2}
\G^{(2)}=\frac{ 9 \l' \, g_2^2}{2 \pi^5} A'_{n_1,n_2}
\]
with
\begin{align}
& A'_{n_1,n_2}
\left\{ 
\begin{aligned}
& =0 \\
&\quad\; \text{for} \quad n_1 = n_2 = 0,\, n_1 = 0 \wedge n_2 \neq 0,\, 
n_1 \neq 0 \wedge n_2 = 0 \, , \, n_1 = - n_2 \notag\\[3mm]
& =\tfrac{1}{32}\,\tfrac{3}{4 n^2}\, \big(\, \tfrac{5}{2} (1+2[Q]) 
- 2 s_{(1)}(1)- 2s_{(1)}(2) + 2s_{(1)}(3) -\tfrac{1}{2}s_{(1)}(4) \,\big)\\
&\quad\; \text{for} \quad n \equiv n_1 = n_2, \, n \equiv n_1 = -2 n_2 \, 
, \, n \equiv -2 n_1 = n_2\\[3mm]
& =\tfrac{1}{32} \tfrac{Q^2}{n_1^2 \, n_2^2 \, (n_1 + n_2 )^2} \big(
 \tfrac{3}{2} (1+2[Q]) - s_{(2)}(n_1) - s_{(2)}(n_2) - s_{(2)}(n_1+n_2)\\
& \qquad\qquad\qquad\quad                      -\tfrac{1}{2} s_{(2)}
( 2 n_1) - \tfrac{1}{2} s_{(2)}(2 n_2) -\tfrac{1}{2} s_{(2)}( 2(n_1+n_2)) \\
& \qquad\qquad\qquad\quad                      + s_{(2)}(n_1 - n_2) 
+ s_{(2)}(2n_1 + n_2) + s_{(2)}(n_1 + 2n_2)\big)\\
& \quad\;\, \text{else,}
\end{aligned} \right.
\end{align}
where we have defined
\begin{equation}\label{sdef}
s_{(1)}(x)= \frac{\sin(\pi \, x \,(1+2[Q])/\sqrt{3})}{\sin(\pi x/\sqrt{3})} 
\quad \text{and} \quad 
s_{(2)}(x)= \frac{\sin(\pi \, x \,(1+2[Q])/Q)}{\sin(\pi x/Q)}.
\end{equation}
\begin{figure} 
\begin{center} 
\includegraphics[height=2.5in]{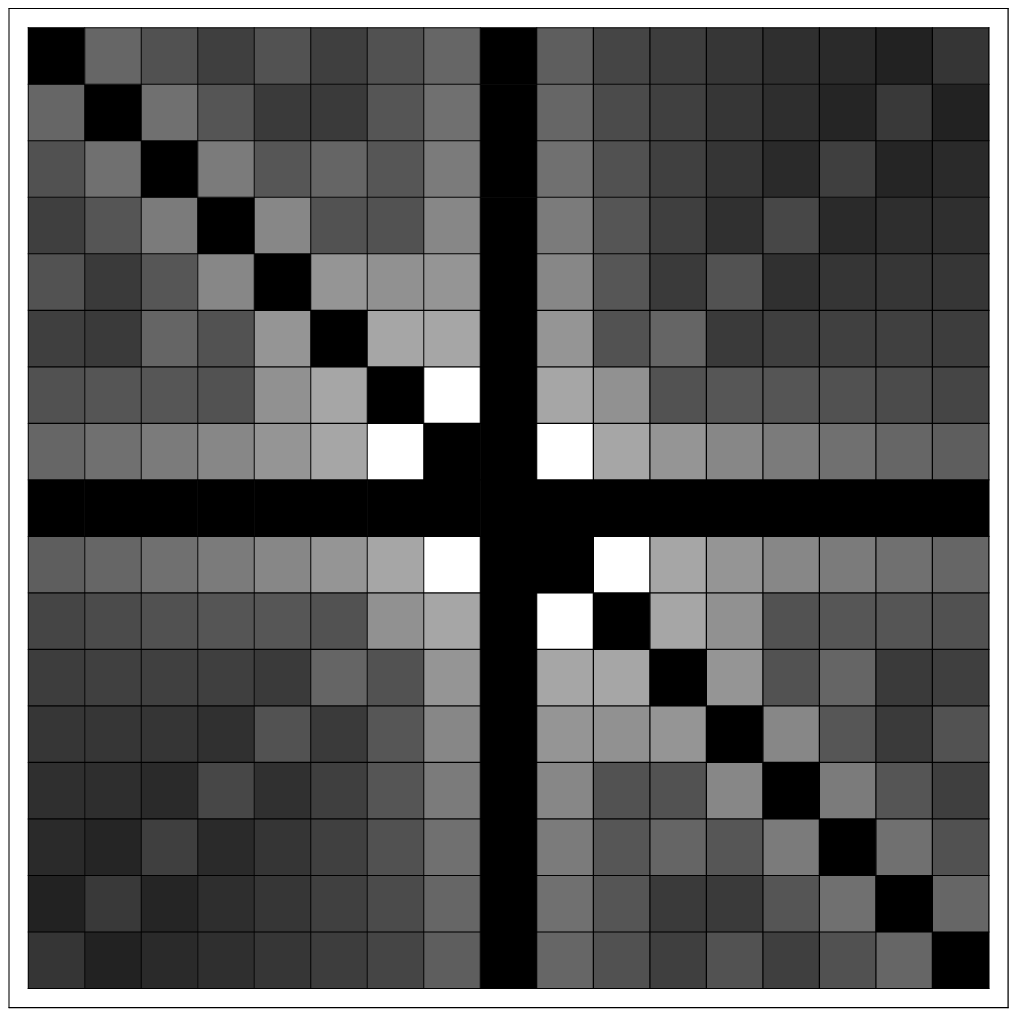} \hspace{5mm}
\includegraphics[height=2.5in]{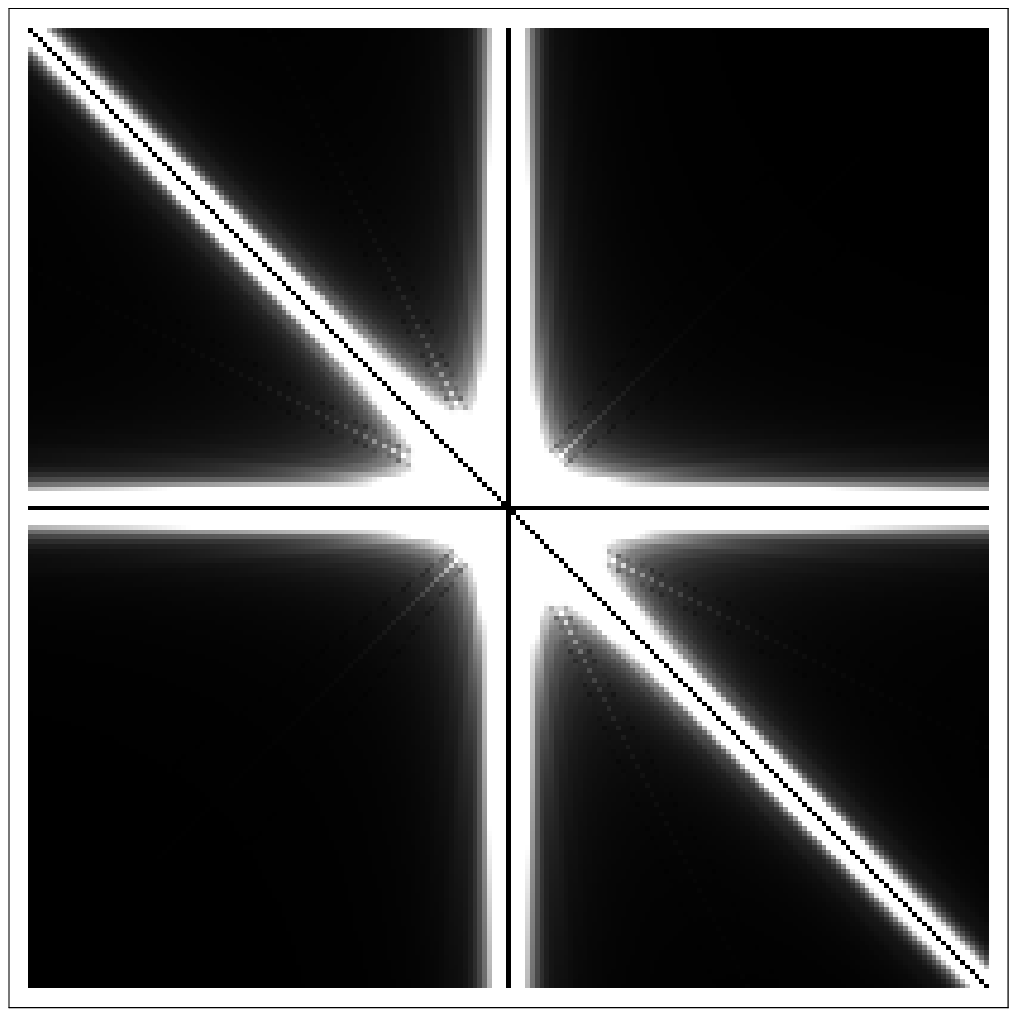}
\end{center}
\caption{The decay width is plotted against $(n_1,n_2)$ (left: decay channel 1 (numerical result),
$n_1,n_2 \in [-8,8]$; right: decay channel 2, $n_1,n_2 \in [-100,100]$), 
where the relative brightness of the cells is a 
measure for the relative size of $\G$ (e.g. $\G=0$ corresponds to black cells).}
\end{figure}
Note that the two classes of symmetry axes reappear in the first two cases
of $A_{n_1,n_2}'$.
In figure 2 the decay width $\G^{(2)}$ is plotted against $n_1$ and $n_2$. The axes
with $E= \l' n^2$ and $E = 3 \l' n^2$ are highly visible. It is
remarkable, that although the range in which $\G^{(1)}$ was computed
is comparatively small, one can already recognize the same structures.

Finally, let us note that the decay width \eqref{decaywidth} could have equally well
been computed with the (non Hermitian!) gauge theory Hamiltonian $H$, a fact
also noted in the two-impurity sector in \cite{GF}. To see this, 
consider the string Hamiltonian matrix element
\[
\langle f|\tilde{H}|i\rangle = \langle f|H|i\rangle +\frac{g_2}{2} (E_i-E_f)\, 
\langle f|\Sigma|i\rangle + {\cal O}(g_2{}^2)
\]
upon using \eqref{Htildedef}. Hence for degenerate states $|i\rangle$ and $|f\rangle$ the
discrepancy between matrix elements of $\tilde{H}$ and $H$ vanishes
up to order $g_2{}^2$ -- provided there are no poles at degeneracy in
$\langle f|\Sigma|i\rangle$, which is not the case. Therefore
the decay width $\Gamma$ of \eqref{decaywidth} may also be written as
\[
\G = \sum_f 2 \pi  \langle i|H|f\rangle\, \langle f|H|i\rangle\,  \d(E_i-E_f)
\]
and the knowledge of the $\Sigma$ matrix elements is irrelevant for the
computation of this physical quantity. 


\section{Conclusions}
\setcounter{equation}{0}

Finally we have a brief look at the 2-loop calculation. In \cite{DilOp} the
2-loop dilatation operator $D_4$ was computed. Applied to two-impurity states
in the BMN limit $D_4$ just acts as the square of $D_2$, i.e.
\[
\lim_{\stackrel{N,J\to \infty,}{{}_{N/J^2\, \text{fixed}}}} 
 {\cal D}_4\circ {\cal O}_{\rm 2 imp} \propto H^2\circ {\cal O}_{\rm 2 imp} \, .
\]
This simple relation does not hold for more than two impurities. Instead, one 
obtains for three impurities at the planar level
\begin{align}
&({D}_4)_{\rm planar}|x_1,x_2,1-x_1-x_2\rangle \propto
\Big[ \p_{x_1}^4 + (\p_{x_1}-\p_{x_2})^4+\p_{x_2}^4\Big]|x_1,x_2,1-x_1-x_2\rangle\;, \
\end{align}
which will lead to ``energy'' eigenvalues $E_{{\mathcal O}(\l'^2)} 
\propto n_1^4 +n_2^4+(n_1+n_2)^4$ not being a perfect square.   
This result confirms what one would expect from the string 
theory point of view from expanding out the square root in \eqref{stringsqr}.

In this paper we have studied the quantum mechanics of BMN gauge theory in the
sector of three scalar impurities. The relevant matrix elements of the Hamiltonian
in the gauge and string theory basis were computed and agreement with 
string field theory was demonstrated. Finally the decay widths of three-impurity 
single trace states into double trace states were found. It was
shown that in the three-impurity sector the instability of single trace
states is a first order process.

Our results represent another non-trivial check of the proposed duality
of plane wave strings and ${\cal N}=4$ Super Yang-Mills. Despite of this
success, a number of worrying disagreements between plane wave string 
and gauge theory are known: 
The reported discrepancy in the two-impurity sector at two quantum-loop order 
($\lambda'^2$) noted in \cite{SV} is unresolved, although there only matrix
elements in string and gauge theory were compared. A priori these need not
agree. It would be highly desirable to compute a physical quantity, such as
an energy shift or a decay width at order $\lambda'{}^2$ in string field theory
to see whether this disagreement is really there.
A further discrepancy consists in the existence of impurity non-conserving 
amplitudes in string field theory at order $\sqrt{\lambda'}$, which obviously 
have no counterpart in perturbative
Yang-Mills \cite{Klebanov}. 
The number of disagreements grows as one moves away from the plane-wave
limit: Incorporating the first curvature corrections to the plane-wave limit 
of superstring theory yields different corrections to the
spectrum than the corresponding $1/J$ corrections on the gauge side,
starting at three loop order in the planar sector
\cite{Callan,nkp}. Similarly, discrepancies at three loop order appear
in the comparisons of gauge theory scaling dimensions and semiclassical
spinning string solutions in the full $AdS_5\times S^5$ theory \cite{spins}.

It remains to be seen how this miraculous mosaic of perturbative
agreements and disagreements of string and gauge theory is to be understood.


\vspace{6mm}
\noindent
{\bf Acknowledgments} \\[.2cm]
\noindent
We would like to thank  Niklas Beisert, Dan Freedman and 
Ari Pankiewicz for interesting 
discussions. 
P.G. thanks the University of Washington for 
hospitality extended to her in the final stages of this 
project.

\appendix
\newpage
\section*{Appendix}\label{appB}
\setcounter{section}{1}
\renewcommand{\thesection}{\Alph{section}}
\renewcommand{\theequation}{\Alph{section}.\arabic{equation}}
\setcounter{equation}{0}

\subsection{Action of the 1-loop vertex operator on ${\mathcal O}_{3;l}$, ${\mathcal O}_{2,1;l}$ and 
${\mathcal O}_{1,1,1;l}$} \label{appH_0}

In this Appendix we present the expressions for $H \circ {\mathcal O}$ (${\mathcal O}$ stands for the three
different types of operators we introduced in section 2), which can be obtained by performing Wick
contractions. We begin with operators of the type ${\mathcal O}_{3;l}$.
The trace-conserving part $H_0 \circ {\mathcal O}_{3;l}$ reads 

\begin{flalign} \label{h0_o(3)}
&{\it H}_0 \circ {\mathcal O}^{\text{\tiny 123}}_{p_1,p_2,J_0-p_1-p_2}
{\mathcal O}_{J_1} \ldots {\mathcal O}_{J_l} = \notag \\[2mm]
&\:\tfrac{g_{\rm YM}^2 N}{8 \pi^2}\Big[
6{\mathcal O}^{\text{\tiny 123}}_{p_1, p_2, J_0-p_1-p_2}
- {\mathcal O}^{\text{\tiny 123}}_{p_1-1, p_2, J_0-\left(p_1-1\right)-p_2} 
- {\mathcal O}^{\text{\tiny 123}}_{p_1+1, p_2, J_0-\left(p_1+1\right)-p_2}\notag\\[2mm]
&\quad\quad\quad
- {\mathcal O}^{\text{\tiny 123}}_{p_1, p_2+1, J_0-p_1-\left(p_2+1\right)}
- {\mathcal O}^{\text{\tiny 123}}_{p_1, p_2-1, J_0-p_1-\left(p_2-1\right)}\notag\\[2mm]
&\quad\quad\quad
- {\mathcal O}^{\text{\tiny 123}}_{p_1+1, p_2-1, J_0-p_1-p_2}
- {\mathcal O}^{\text{\tiny 123}}_{p_1-1, p_2+1, J_0-p_1-p_2}\Big]
{\mathcal O}_{J_1}\ldots{\mathcal O}_{J_l} \, ,\hspace{1cm}
\end{flalign}

\vspace{4mm}
\noindent
while $H_{\pm} \circ {\mathcal O}_{3;l}$, where the number of traces in- and decreases, are given by   

\begin{flalign}
&{\it H}_+ \circ {\mathcal O}^{\text{\tiny 123}}_{p_1,p_2,J_0-p_1-p_2}
{\mathcal O}_{J_1} \ldots {\mathcal O}_{J_l} = \notag\\
&\: \tfrac{g_{\rm YM}^2}{8 \pi^2} \bigg\{ 
\sum_{i=1}^{p_1-1} 
\Big(2 {\mathcal O}^{\text{\tiny 123}}_{p_1-i, p_2, J_0-p_1-p_2}
- {\mathcal O}^{\text{\tiny 123}}_{p_1-i-1, p_2+1, J_0-p_1-p_2}\notag\\
&\qquad\qquad\quad
- {\mathcal O}^{\text{\tiny 123}}_{p_1-i-1, p_2, J_0-\left(p_1-1\right)-p_2} \Big)
{\mathcal O}_i\notag\\[-1mm]
&\quad\: 
+ \sum_{i=1}^{p_2-1} 
\Big(2 {\mathcal O}^{\text{\tiny 123}}_{p_1, p_2-i, J_0-p_1-p_2}
- {\mathcal O}^{\text{\tiny 123}}_{p_1, p_2-i-1, J_0-p_1-\left(p_2-1\right)}\notag\\
&\qquad\qquad\quad
- {\mathcal O}^{\text{\tiny 123}}_{p_1+1, p_2-i-1, J_0-p_1-p_2}\Big)
{\mathcal O}_i\notag\\[-1mm]
&\quad\: 
+\hspace{-2mm}  \sum_{i=1}^{J_0-p_1-p_2-1} \hspace{-2mm}
\Big(2 {\mathcal O}^{\text{\tiny 123}}_{p_1, p_2, J_0-p_1-p_2-i}
- {\mathcal O}^{\text{\tiny 123}}_{p_1+1, p_2, J_0-\left(p_1+1\right)-p_2-i}\notag\\
&\qquad\qquad\quad 
- {\mathcal O}^{\text{\tiny 123}}_{p_1, p_2+1, J_0-p_1-\left(p_2+1\right)-i} \Big)
{\mathcal O}_i \notag\\
&\quad\: +
\sum_{i=0}^{p_1-1}
\Big( {\mathcal O}^{\text{\tiny 1}}_{J_0-p_1-p_2+i}
\Big[ {\mathcal O}^{\text{\tiny 23}}_{p_2,p_1-i}
- {\mathcal O}^{\text{\tiny 23}}_{p_2+1,p_1-i-1}\Big]\notag\\
&\qquad\qquad\quad  
+ {\mathcal O}^{\text{\tiny 2}}_{p_2+i}
\Big[ {\mathcal O}^{\text{\tiny 31}}_{J_0-p_1-p_2,p_1-i}
- {\mathcal O}^{\text{\tiny 31}}_{J_0-\left(p_1-1\right)-p_2,p_1-i-1}\Big]\Big)\notag \hspace{2cm}
\end{flalign}

\begin{flalign}
&\quad\: +\sum_{i=0}^{p_2-1}
\Big( {\mathcal O}^{\text{\tiny 2}}_{p_1+i}
\Big[ {\mathcal O}^{\text{\tiny 31}}_{J_0-p_1-p_2,p_2-i}
- {\mathcal O}^{\text{\tiny 31}}_{J_0-p_1-\left(p_2-1\right),p_2-i-1}\Big]\notag\\
&\qquad\qquad\quad 
+ {\mathcal O}^{\text{\tiny 3}}_{J_0-p_1-p_2+i}
\Big[ {\mathcal O}^{\text{\tiny 12}}_{p_1,p_2-i}
- {\mathcal O}^{\text{\tiny 12}}_{p_1+1,p_2-i-1}\Big]\Big)\notag\\[-1mm]
&\quad\: 
+ \hspace{-2mm} \sum_{i=0}^{J_0-p_1-p_2-1} \hspace{-2mm} 
\Big( {\mathcal O}^{\text{\tiny 3}}_{p_2+i}
\Big[ {\mathcal O}^{\text{\tiny 12}}_{p_1,J_0-p_1-p_2-i}
- {\mathcal O}^{\text{\tiny 12}}_{p_1+1,J_0- \left(p_1+1 \right)-p_2-i}\Big]\notag\\[-1mm]
&\qquad\qquad\quad
+ {\mathcal O}^{\text{\tiny 1}}_{p_1+i}
\Big[ {\mathcal O}^{\text{\tiny 23}}_{p_2,J_0-p_1-p_2-i}
- {\mathcal O}^{\text{\tiny 23}}_{p_2+1,J_0-p_1-\left(p_2+1\right)-i}\Big] \Big)\bigg\} 
{\mathcal O}_{J_1} \ldots {\mathcal O}_{J_l} \hspace{0.7cm}
\end{flalign}

\vspace{-2mm}
\noindent
and

\begin{flalign}
&{\it H}_- \circ {\mathcal O}^{\text{\tiny 123}}_{p_1,p_2,J_0-p_1-p_2}
{\mathcal O}_{J_1} \ldots {\mathcal O}_{J_l} = \notag\\
&\: \tfrac{g_{\rm YM}^2}{8 \pi^2}
\sum_{i=1}^{l} J_i
\Big( 2 {\mathcal O}^{\text{\tiny 123}}_{p_1+J_i, p_2, J_0-p_1-p_2}
- {\mathcal O}^{\text{\tiny 123}}_{p_1+J_i-1, p_2+1, J_0-p_1-p_2}\notag\\[-1mm]
&\qquad\qquad\quad  
- {\mathcal O}^{\text{\tiny 123}}_{p_1+J_i-1, p_2, J_0-\left(p_1-1\right)-p_2}\notag\\[2mm]
&\qquad\qquad\quad
+2 {\mathcal O}^{\text{\tiny 123}}_{p_1, p_2+J_i, J_0-p_1-p_2}
- {\mathcal O}^{\text{\tiny 123}}_{p_1, p_2+J_i-1, J_0-p_1-\left(p_2-1\right)}\notag\\[2mm]
&\qquad\qquad\quad  
- {\mathcal O}^{\text{\tiny 123}}_{p_1+1, p_2+J_i-1, J_0-p_1-p_2}\notag\\[2mm]
&\qquad\qquad\quad
+2 {\mathcal O}^{\text{\tiny 123}}_{p_1, p_2, J_0-p_1-p_2+J_i}
-  {\mathcal O}^{\text{\tiny 123}}_{p_1+1, p_2, J_0-\left(p_1+1\right)-p_2+J_i}\notag\\[2mm]
&\qquad\qquad\quad 
- {\mathcal O}^{\text{\tiny 123}}_{p_1, p_2+1, J_0-p_1-\left(p_2+1\right)+J_i}\Big)
\times{\mathcal O}_{J_1} \ldots \makebox[0pt]{$\not$}{\mathcal O}_{J_i} \ldots {\mathcal O}_{J_l}\,.
\hspace{1.5cm}
\end{flalign}

\vspace{3mm}
\noindent
The same can be done for ${\mathcal O}_{2,1;l}$ with

\begin{flalign} \label{h0_o(2,1)}
&{\it H}_0 \circ {\mathcal O}^{\text{\tiny ab}}_{p,J_0-p-q}{\mathcal O}^{\text{\tiny c}}_{q}
{\mathcal O}_{J_1} \ldots {\mathcal O}_{J_l} = \notag\\[2mm]
&\;\tfrac{g_{\rm YM}^2 N}{4 \pi^2}\Big[2{\mathcal O}^{\text{\tiny ab}}_{p,J_0-p-q}
-{\mathcal O}^{\text{\tiny ab}}_{p+1,J_0-(p+1)-q}-{\mathcal O}^{\text{\tiny ab}}_{p-1,J_0-(p-1)-q}\Big]
{\mathcal O}^{\text{\tiny c}}_{q}{\mathcal O}_{J_1} \ldots {\mathcal O}_{J_l}\, ,\\[3mm]
%
&{\it H}_+ \circ {\mathcal O}^{\text{\tiny ab}}_{p,J_0-p-q}{\mathcal O}^{\text{\tiny c}}_{q}
{\mathcal O}_{J_1} \ldots {\mathcal O}_{J_l} = \notag\\
&\;\tfrac{g_{\rm YM}^2}{4 \pi^2}\bigg[
\sum_{i=1}^{p-1}\Big({\mathcal O}^{\text{\tiny ab}}_{p-i,J_0-p-q}
-{\mathcal O}^{\text{\tiny ab}}_{p-1-i,J_0-(p-1)-q}\Big){\mathcal O}_{i}\notag\\
&\quad 
+ \hspace{-2mm} \sum_{i=1}^{J_0-p-q-1} \hspace{-2mm} 
\Big({\mathcal O}^{\text{\tiny ab}}_{p,J_0-p-q-i}
-{\mathcal O}^{\text{\tiny ab}}_{p+1,J_0-(p+1)-q-i}\Big){\mathcal O}_{i}\bigg]
{\mathcal O}^{\text{\tiny c}}_{q}{\mathcal O}_{J_1} \ldots {\mathcal O}_{J_l}
\end{flalign}

\noindent
and

\begin{flalign}
&{\it H}_- \circ {\mathcal O}^{\text{\tiny ab}}_{p,J_0-p-q}{\mathcal O}^{\text{\tiny c}}_{q}
{\mathcal O}_{J_1} \ldots {\mathcal O}_{J_l} = \notag\\
&\;\tfrac{g_{\rm YM}^2}{4 \pi^2} \bigg\{
\sum_{i=1}^{l}J_i\Big[
{\mathcal O}^{\text{\tiny ab}}_{p+J_i,J_0-p-q}-{\mathcal O}^{\text{\tiny ab}}_{p-1+J_i,J_0-(p-1)-q}\notag\\
&\qquad\qquad\quad 
+{\mathcal O}^{\text{\tiny ab}}_{p,J_0-p-q+J_i}-{\mathcal O}^{\text{\tiny ab}}_{p+1,J_0-(p+1)-q+J_i}\Big]
{\mathcal O}^{\text{\tiny c}}_{q}{\mathcal O}_{J_1} \ldots \makebox[0pt]{$\not$}{\mathcal O}_{J_i} 
\ldots {\mathcal O}_{J_l} \notag\\
&\quad+
\tfrac{1}{2}\bigg[
\sum_{i=0}^{p-1}
\Big({\mathcal O}^{\text{\tiny acb}}_{q+i,p-i,J_0-p-q}
-{\mathcal O}^{\text{\tiny acb}}_{q+1+i,p-1-i,J_0-p-q}
\notag\\
&\qquad\qquad\quad  
+ {\mathcal O}^{\text{\tiny acb}}_{p-i,q+i,J_0-p-q}
-{\mathcal O}^{\text{\tiny acb}}_{p-1-i,q+1+i,J_0-p-q}
\Big) \notag\\
&\qquad\; +  
\hspace{-2mm} \sum_{i=0}^{J_0-p-q-1}\hspace{-2mm}
\Big({\mathcal O}^{\text{\tiny abc}}_{p,q+i,J_0-p-q-i}
-{\mathcal O}^{\text{\tiny abc}}_{p,q+1+i,J_0-p-(q+1)-i}\notag\\
&\qquad\qquad\quad 
+ {\mathcal O}^{\text{\tiny abc}}_{p,J_0-p-q-i,q+i}
-{\mathcal O}^{\text{\tiny abc}}_{p,J_0-p-(q+1)-i,q+1+i}\Big)\notag\\
&\qquad\; +
\sum_{i=0}^{q-1}\Big(
{\mathcal O}^{\text{\tiny abc}}_{p,q-i,J_0-p-q+i}
-{\mathcal O}^{\text{\tiny abc}}_{p+1,q-1-i,J_0-p-q+i}\notag\\
&\qquad\qquad\quad 
+ {\mathcal O}^{\text{\tiny abc}}_{p,J_0-p-q+i,q-i}
-{\mathcal O}^{\text{\tiny abc}}_{p+1,J_0-p-q+i,q-1-i}\notag\\[2mm]
&\qquad\qquad\quad
+ {\mathcal O}^{\text{\tiny acb}}_{q-i,p+i,J_0-p-q}
-{\mathcal O}^{\text{\tiny acb}}_{q-1-i,p+i,J_0-p-(q-1)}\notag\\
&\qquad\qquad\quad 
+ {\mathcal O}^{\text{\tiny acb}}_{p+i,q-i,J_0-p-q}
-{\mathcal O}^{\text{\tiny acb}}_{p+i,q-1-i,J_0-p-(q-1)}\Big)\bigg]
{\mathcal O}_{J_1} \ldots {\mathcal O}_{J_l}\bigg\}\,.
\hspace{0.2cm}
\end{flalign}

\vspace{3mm}
\noindent
The only non-vanishing contribution of $H \circ {\mathcal O}_{1,1,1;l}$ is

\begin{flalign}
&{\it H}_- \circ {\mathcal O}^{\text{\tiny 1}}_{p_1}{\mathcal O}^{\text{\tiny 2}}_{p_2}
{\mathcal O}^{\text{\tiny 3}}_{J_0-p_1-p_2}{\mathcal O}_{J_1} \ldots {\mathcal O}_{J_l}= \notag\\
&\;\tfrac{g_{\rm YM}^2}{8 \pi^2}\bigg[
\sum_{i=0}^{p_1-1}
\Big(\Big[
{\mathcal O}^{\text{\tiny 12}}_{p_2+i,p_1-i}
-{\mathcal O}^{\text{\tiny 12}}_{p_2+1+i,p_1-1-i}\notag\\[-1.5mm]
&\qquad\qquad\quad
+ {\mathcal O}^{\text{\tiny 12}}_{p_1-i,p_2+i}
- {\mathcal O}^{\text{\tiny 12}}_{p_1-1-i,p_2+1+i}\Big]
{\mathcal O}^{\text{\tiny 3}}_{J_0-p_1-p_2} \notag\\
&\qquad\qquad +
\Big[{\mathcal O}^{\text{\tiny 13}}_{J_0-p_1-p_2+i,p_1-i}
- {\mathcal O}^{\text{\tiny 13}}_{J_0-(p_1-1)-p_2+i,p_1-1-i}\notag\\
&\qquad\qquad\quad +
{\mathcal O}^{\text{\tiny 13}}_{p_1-i,J_0-p_1-p_2+i}
- {\mathcal O}^{\text{\tiny 13}}_{p_1-1-i,J_0-(p_1-1)-p_2+i}\Big]
{\mathcal O}^{\text{\tiny 2}}_{p_2}\Big) \notag
\hspace{2cm}
\end{flalign}

\begin{flalign}
&\quad\;+
\sum_{i=0}^{p_2-1}
\Big(\Big[{\mathcal O}^{\text{\tiny 12}}_{p_2-i,p_1+i}
-{\mathcal O}^{\text{\tiny 12}}_{p_2-1-i,p_1+1+i}\notag\\[-1.5mm]
&\qquad\qquad\quad
+ {\mathcal O}^{\text{\tiny 12}}_{p_1+i,p_2-i}
- {\mathcal O}^{\text{\tiny 12}}_{p_1+1+i,p_2-1-i}\Big]
{\mathcal O}^{\text{\tiny 3}}_{J_0-p_1-p_2}\notag\\[-1.5mm]
&\qquad\qquad +
\Big[{\mathcal O}^{\text{\tiny 23}}_{J_0-p_1-p_2+i,p_2-i}
- {\mathcal O}^{\text{\tiny 23}}_{J_0-p_1-(p_2-1)+i,p_2-1-i}\notag\\
&\qquad\qquad\quad +
{\mathcal O}^{\text{\tiny 23}}_{p_2-i,J_0-p_1-p_2+i}
- {\mathcal O}^{\text{\tiny 23}}_{p_2-1-i,J_0-p_1-(p_2-1)+i}\Big]
{\mathcal O}^{\text{\tiny 1}}_{p_1}\Big)\notag\\[-1.5mm]
&\quad\;+
\hspace{-2mm} \sum_{i=0}^{J_0-p_1-p_2-1} \hspace{-2mm}
\Big(\Big[{\mathcal O}^{\text{\tiny 13}}_{J_0-p_1-p_2-i,p_1+i}
- {\mathcal O}^{\text{\tiny 13}}_{J_0-(p_1+1)-p_2-i,p_1+1+i}\notag\\[-1.5mm]
&\qquad\qquad\quad 
+ {\mathcal O}^{\text{\tiny 13}}_{p_1+i,J_0-p_1-p_2-i}
- {\mathcal O}^{\text{\tiny 13}}_{p_1+1+i,J_0-(p_1+1)-p_2-i}\Big]{\mathcal O}^{\text{\tiny 2}}_{p_2}\notag\\
&\qquad\qquad+
\Big[{\mathcal O}^{\text{\tiny 23}}_{J_0-p_1-p_2-i,p_2+i}
- {\mathcal O}^{\text{\tiny 23}}_{J_0-p_1-(p_2+1)-i,p_2+1+i}\notag\\
&\qquad\qquad\quad
+ {\mathcal O}^{\text{\tiny 23}}_{p_2+i,J_0-p_1-p_2-i}
- {\mathcal O}^{\text{\tiny 23}}_{p_2+1+i,J_0-p_1-(p_2+1)-i}\Big]
{\mathcal O}^{\text{\tiny 1}}_{p_1}\Big)\bigg]
{\mathcal O}_{J_1} \ldots {\mathcal O}_{J_l}
\end{flalign}


\subsection{${H}_-$ and $\S_-$} \label{H&S}

Here we summarize the action of $H_-$ and $\Sigma _-$ on the different types of continuum states. 
The action of $H_-$ can be deduced out of the corresponding expressions of Appendix A.1 
by taking $J \to \infty$. One gets 
 
\begin{flalign}
&{\it {H}}_- |n_1, n_2; r_0\rangle|r_1\rangle \ldots |r_l \rangle =\notag\\
&\; \frac{\l'}{2 \pi^3}\sum_{i=1}^l \sum_{m_1,m_2}\bigg(
\frac{\sqrt{r_i}}{r_0}\sin (\pi m_1 \tfrac{r_0}{r_0+r_i})
\sin (\pi m_2 \tfrac{r_0}{r_0+r_i})\sin (\pi (m_1+m_2) \tfrac{r_0}{r_0+r_i})\notag\\
&\qquad \times
\bigg[\frac{n_1(-m_1-2 m_2)}{m_1-n_1\tfrac{r_0+r_{i}}{r_0}}
+ \frac{n_2(2m_1+ m_2)}{m_2-n_2\tfrac{r_0+r_{i}}{r_0}}
+ \frac{-(n_1+n_2)(m_2-m_1)}{-(m_1+m_2)+(n_1+n_2)\tfrac{r_0+r_{i}}{r_0}}\bigg]\notag\\
&\qquad \times
\frac{1}{m_1 n_2-m_2 n_1}\bigg)\;|m_1, m_2; r_0+r_{i}\rangle 
|r_1\rangle \ldots\makebox[0pt]{\;\;$\not$}|r_i\rangle \ldots |r_{l}\rangle
\end{flalign}

\pagebreak
\noindent
and

\begin{flalign}
&{\it {H}}_-|n;r_0-s\rangle^{\text{\tiny ab}}|s\rangle^{\text{\tiny c}}|r_1\rangle \ldots |r_l \rangle=
\notag\\
&\;\frac{\l'}{\pi^2}\bigg( \sum_{i=1}^l\sum_m 
\sqrt{\frac{r_i}{(r_0-s+r_i)(r_0-s)}}\notag\\
&\qquad\quad\times
\frac{m \sin^2(\pi m \frac{r_0-s}{r_0-s+r_i})}{m-n\tfrac{r_0-s+r_i}{r_0-s}}\;
|m;r_0-s+r_i\rangle^{\text{\tiny ab}}|s\rangle^{\text{\tiny c}}
|r_1\rangle \ldots \makebox[0pt]{\;\;$\not$}|r_i\rangle\ldots |r_l\rangle\notag\\
&\quad+
\frac{1}{2}\sum_{m_1,m_2}\frac{\sqrt{r_0-s}}{r_0\,\pi} 
\sin (\pi m_1 \tfrac{s}{r_0})\sin (\pi m_2 \tfrac{s}{r_0})\sin (\pi (m_1+m_2) 
\tfrac{s}{r_0})\notag\\
&\qquad\quad\times
\sum_{(a,b)(c)}\frac{1}{M_c}\Big(\frac{M_c-M_b}{n+M_a\tfrac{r_0-s}{r_0}}
-\frac{M_c-M_a}{n-M_b\tfrac{r_0-s}{r_0}}\Big)|m_1,m_2;r_0\rangle|r_1\rangle \ldots |r_l \rangle
\bigg)
\hspace{2cm}
\end{flalign}

\noindent
as well as 

\begin{flalign}
&{\it {H}}_-|t_1\rangle^{\text{\tiny 1}}|t_2\rangle^{\text{\tiny 2}}|t_3\rangle^{\text{\tiny 3}}
|r_1\rangle \ldots |r_{l}\rangle=\notag\\
&\; -\frac{\l'}{\pi^2}\sum_m \sum_{(a,b)(c)} \frac{\sin^2(\pi m \tfrac{t_a}{t_a+t_b})}{\sqrt{t_a+t_b}}\;
|m; t_a+t_b\rangle^{\text{\tiny ab}}|t_c\rangle^{\text{\tiny c}}|r_1\rangle \ldots |r_l \rangle\,.
\hspace{1.5cm}
\end{flalign}

\noindent
On the other hand, the matrix 
elements of $\Sigma_-$ can be read off from the corresponding two-point 
functions, where again $J$ is taken to infinity. Then we obtain for the action on states like 
$|n_1,n_2;r_0\rangle \ldots$     

\begin{flalign}
&\Sigma_- |n_1,n_2;r_0\rangle |r_1\rangle \ldots |r_l\rangle\notag\\
&\:= -\sum_{i=1}^l \sum_{m_1,m_2} 
\frac{\sin(\pi m_1\tfrac{r_0}{r_0+r_{i}}) \, \sin(\pi m_2\tfrac{r_0}{r_0+r_{i}}) \,
\sin(\pi\left(m_1+m_2\right)\tfrac{r_0}{r_0+r_{i}})}
{(m_1-n_1\tfrac{r_0+r_i}{r_0})(m_2-n_2\tfrac{r_0+r_i}{r_0})(-(m_1+m_2)+(n_1+n_2)\tfrac{r_0+r_i}{r_0})}
\notag\\
&\quad\quad\quad\quad \times 
\frac{\left(r_0+r_i\right)^2\sqrt r_i}{r_0 \pi^3}\;|m_1,m_2;r_0+r_i\rangle 
|r_1\rangle \ldots \makebox[0pt]{\;\;$\not$}|r_i\rangle\ldots |r_l\rangle \notag\\ 
&\quad+
\frac{1}{2}\sum_{\substack{i,j=0 \\ i \neq j}}^l \sqrt{r_i r_j (r_i+r_j)}\;
|n_1,n_2;r_0\rangle |r_1\rangle \ldots\makebox[0pt]{\;\;$\not$}|r_i\rangle 
\ldots \makebox[0pt]{\;\;$\not$}|r_j\rangle\ldots |r_l\rangle|r_i+r_j\rangle \,,
\end{flalign}

\noindent
for states like $|n;r_0-s\rangle^{\text{\tiny ab}}|s\rangle^{\text{\tiny c}} \ldots$

\begin{flalign}
&\Sigma_- |n;r_0-s\rangle^{\text{\tiny ab}}|s\rangle^{\text{\tiny c}}
|r_1\rangle \ldots |r_{l}\rangle \notag\\
&\:= \sum_{i=1}^l \Big( 
\sum_m \sqrt{ \frac{r_i(r_0-s+r_i)^3}{r_0-s}}
\frac{\sin^2(\pi m \tfrac{r_0-s}{r_0-s+r_i})}{\pi^2(m-n \tfrac{r_0-s+r_i}{r_0-s})^2}\;
|m;r_0-s+r_i\rangle^{\text{\tiny ab}}|s\rangle^{\text{\tiny c}}\notag\\
&\quad\quad\quad\quad +
\sqrt{r_i}\; s\;|n;r_0-s\rangle^{\text{\tiny ab}}|s+r_i\rangle^{\text{\tiny c}} \Big)
|r_1\rangle \ldots \makebox[0pt]{\;\;$\not$}|r_i\rangle\ldots |r_l\rangle\notag\\
&\quad+
\frac{1}{2}\sum_{\substack{i,j=0 \\ i \neq j}}^l \;\sqrt{r_i r_j (r_i+r_j)}\;
|n;r_0-s\rangle^{\text{\tiny ab}}|s\rangle^{\text{\tiny c}}
|r_1\rangle \ldots\makebox[0pt]{\;\;$\not$}|r_i\rangle \ldots \makebox[0pt]{\;\;$\not$}|r_j\rangle
\ldots |r_l\rangle|r_i+r_j\rangle\notag\\ 
&\quad -
\sum_{m_1,m_2} \frac{r_0^2}{\pi^3 (r_0-s)^{1/2}}\;\sin(\pi m_1 \tfrac{s}{r_0}) \sin(\pi m_2 \tfrac{s}{r_0})
 \sin(\pi \left(m_1+m_2\right) \tfrac{s}{r_0})\notag\\
&\quad\quad\quad \times
\left(\left(M_a+n\tfrac{r_0}{r_0-s}\right)\left(M_b-n\tfrac{r_0}{r_0-s}\right)M_c\right)^{-1}\;
|m_1,m_2;r_0\rangle |r_1\rangle \ldots |r_l\rangle
\end{flalign}

\noindent
and finally for states like 
$|t_1\rangle^{\text{\tiny 1}}|t_2\rangle^{\text{\tiny 2}}|t_3\rangle^{\text{\tiny 3}}\ldots$

\begin{flalign}
&\Sigma_-|t_1\rangle^{\text{\tiny 1}}|t_2\rangle^{\text{\tiny 2}}|t_3\rangle^{\text{\tiny 3}}
|r_1\rangle \ldots |r_{l}\rangle \notag\\
&\;=\sum_{i=1}^l \sqrt{r_i}\; \Big(
t_1 \; |t_1+r_i\rangle^{\text{\tiny 1}}|t_2\rangle^{\text{\tiny 2}}|t_3\rangle^{\text{\tiny 3}}
+ t_2 \; |t_1\rangle^{\text{\tiny 1}}|t_2+r_i\rangle^{\text{\tiny 2}}|t_3\rangle^{\text{\tiny 3}}\notag\\
&\qquad\qquad
+ t_3 \; |t_1\rangle^{\text{\tiny 1}}|t_2\rangle^{\text{\tiny 2}}|t_3+r_i\rangle^{\text{\tiny 3}}\Big)|r_1\rangle... \makebox[0pt]{\;\;$\not$}|r_i\rangle...|r_l\rangle\notag\\
&\quad+
\frac{1}{2} \sum_{\substack{i,j=0 \\ i \neq j}}^l \; \sqrt{r_i r_j (r_i+r_j)} \; 
|t_1\rangle^{\text{\tiny 1}}|t_2\rangle^{\text{\tiny 2}}|t_3\rangle^{\text{\tiny 3}}
|r_1\rangle \ldots \makebox[0pt]{\;\;$\not$}|r_i\rangle \ldots \makebox[0pt]{\;\;$\not$}|r_j\rangle
\ldots |r_l\rangle|r_i+r_j\rangle\notag\\
&\quad-
\sum_m \sum_{(a,b)(c)}\frac{(t_a+t_b)^{3/2}}{m^2\; \pi^2} \; 
\sin^2(\pi m \tfrac{t_a}{t_a+t_b}) \; 
|m;t_a+t_b\rangle^{\text{\tiny ab}}|t_c\rangle^{\text{\tiny c}}|r_1\rangle \ldots |r_{l}\rangle\, ,
\end{flalign}

with $t_1+t_2+t_3=r_0$ and $M_1:=-(m_1+m_2)$,\;$M_2:=m_1$,\;$M_3:=m_2$.\\


\section{The sum $A_{n_1,n_2}$}
\setcounter{equation}{0}

In the following we will use the short cuts
\begin{equation}
 p \equiv \frac{\pi \, n_1}{Q} \quad \text{and} \quad q \equiv 
\frac{\pi \, n_2}{Q} \, .
\end{equation}
The sum $A_{n_1,n_2}$ then reads
\begin{align}
&   A_{p,q}= \sum_{m=1}^{[Q]}\, \sin^2 (p \, m) \, \sin^2 (q \, m) \, 
\sin^2 ((p+q)\,m) \\[-1mm]
& \qquad  =\tfrac{1}{32} \sum_{m=1}^{[Q]}\, 
                     \big(3 - 2 \cos (2 \, p \, m) - 2 \cos (2 \, q \, m) 
- 2 \cos (2 \, (p+q) \, m) \notag\\[-2mm] 
& \qquad\qquad\quad\: - \;\cos (4 \, p \, m) - \;\cos (4 \, q \, m) - \;
\cos (4 \, (p+q) \, m)     \notag\\[2mm]
& \qquad\qquad\quad\: + 2 \cos (2 \, (p-q) \, m) + 2 \cos (2 \, (2p+q) \, m) 
+ 2 \cos (2 \, (p+2q) \, m) \big).
\nn
\end{align}
Before calculating the sum we study the cases, where at least one of the 
arguments of the cosines vanishes, namely
\begin{itemize}
  \item $p = q = 0$, $p = 0 \wedge q \neq 0$, $p \neq 0 \wedge q = 0$ and $p = - q$ 
\, : \, $\G^{(2)} \rightarrow 0$
  \item $r\equiv p = q$, $r \equiv p = -2 q $ and $r \equiv -2 p = q$\; :\\[2mm]
	$\quad A_{p,q} \rightarrow A_r = \tfrac{1}{32} \sum_{m=1}^{[Q]} 
		\big(5 - 4 \cos (2 \, r \,  m) - 4 \cos (4 \, r \,  m) + 4 
\cos (6 \, r \,  m) - \cos (8 \, r \,  m)\big)$\,.
\nn
\end{itemize} 
Using 
\begin{equation}
\sum_{m=1}^{[Q]} \cos (x\,m) = - 1 + \cos(\tfrac{[Q]\,x}{2}) \csc(\tfrac{x}{2})
 \sin(\tfrac{x}{2}(1+[Q])) \quad \text{for} \quad \frac{x}{2 \pi} \not\in \mathbb{Z}
\nn
\end{equation} 
$A_{n_1,n_2}$ as stated in \eqref{gamma2} and \eqref{sdef} follows.

\bigskip



\end{document}